\documentclass[%
reprint,
superscriptaddress,
nofootinbib,
showkeys,
showpacs,
amsmath,amssymb,
aps,
]{revtex4-1}

\usepackage{graphicx,color}
\usepackage{dcolumn}
\usepackage{url} 
\usepackage{amsmath,amsfonts,amsthm,bm,amssymb} 
\usepackage{booktabs}
\graphicspath{ {figures/} }
\usepackage{hyperref,comment}
\usepackage{footmisc}
\usepackage{float}
\usepackage{soul}
\usepackage{natbib}
\setcitestyle{numbers,sort&compress}

\newcommand{\R}{{\cal R}}
\newcommand{\I}{{\cal I}}
\newcommand{\re}{{\rm Re}}

\let\a=\alpha \let\b=\beta  \let\g=\gamma   \let\e=\varepsilon
      \let\k=\kappa 
\let\m=\mu    \let\n=\nu         \let\p=\pi    
 \let\t=\tau    
     
\let\G=\Gamma \let\D=\Delta  \let\L=\Lambda

\font\tenmib=cmmib10\font\sevenmib=cmmib7\font\fivemib=cmmib5%
\def\BDpr {{\mbox{\boldmath$ \partial$}}}

\def\eqalign#1{\null\,\vcenter{\openup\jot
		\ialign{\strut\hfil$\displaystyle{##}$&$\displaystyle{{}##}$\hfil
			\crcr#1\crcr}}\,}

\def\EE{{\mathcal E}}\def\DD{{\mathcal D}}

\def\uu{{\boldsymbol{u}}}
\def\kk{{\V k}}
\def\pp{{\V p}}
\def\qq{{\V q}}
\def\xx{{\V x}}
\def\ff{{\V f}}
\def\nn{{\bf n}}
\def\T#1{{#1_{\kern-3pt\lower7pt\hbox{$\widetilde{}$}}\kern3pt}}

\def\ie{i.e.}

\def\defi{{\buildrel def\over=}}

\def\otto{\,{\kern-1.truept\leftarrow\kern-5.truept\to\kern-1.truept}\,}
\def\DD{{\cal D}}
\newdimen\xshift \newdimen\xwidth \newdimen\yshift \newdimen\ywidth

\def\ins#1#2#3{\vbox to0pt{\kern-#2pt\hbox{\kern#1pt #3}\vss}\nointerlineskip}

\def\eqfig#1#2#3#4#5{
	\par\xwidth=#1pt \xshift=\hsize \advance\xshift
	by-\xwidth \divide\xshift by 2
	\yshift=#2pt \divide\yshift by 2
	{\hglue\xshift \vbox to #2pt{\vfil
			#3 \includegraphics{#4.eps}
		}\hfill\raise\yshift\hbox{#5}}}

\def\V#1{{\boldsymbol{#1}}}
\def\lis#1{{\overline#1}}

\def\tende#1{\,\vtop{\ialign{##\crcr\rightarrowfill\crcr
			\noalign{\kern-1pt\nointerlineskip} \hskip3.pt${\scriptstyle
				#1}$\hskip3.pt\crcr}}\,}
\def\eg{e.g.}

\def\0{\noindent}\def\ig{\int}
\newbox\strutboxa
\setbox\strutboxa=\hbox{\vrule height8.5pt depth2.25pt width0pt}
\def\struta{\relax\ifmmode\copy\strutboxa\else\unhcopy\strutboxa\fi}
\def\W#1{#1_{\kern-3pt\lower7.5pt\hbox{$\widetilde{}$}}\kern2pt\,\struta}
\def\*{\vskip2mm}
\def\media#1{\langle #1 \rangle}
\def\Eq#1{\label{#1}}
\def\equ#1{(\ref{#1})}
\def\eE{{\bf e}}

\newcounter{appendice}

\def\be{\begin{equation}}%
\def\ee{\end{equation}}%

\definecolor{iblue}{RGB}{65,105,225}
\definecolor{ired}{RGB}{220,20,60}
\definecolor{igreen}{RGB}{50,205,50}
\definecolor{ipurple}{RGB}{75,0,130}
\definecolor{iochre}{RGB}{218,165,32}
\definecolor{iteal}{RGB}{51,204,204} 
\definecolor{imauve}{RGB}{204,51,153}

\begin{document}

	\title{Non-equilibrium Ensembles for the three-dimensional Navier-Stokes equations}

	\author{G. Margazoglou}
	\email[Present address: Department of Aeronautics, Imperial College London, South Kensington Campus, London SW7 2AZ, United Kingdom; Corresponding author:]{g.margazoglou@imperial.ac.uk}
	\affiliation{Department of Mathematics and Statistics, University of Reading, Reading, United Kingdom}
	\affiliation{Centre for the Mathematics of Planet Earth, University of Reading, Reading, United Kingdom}

	\author{L.\ Biferale}
	\affiliation{Department of Physics and INFN, University of Rome Tor Vergata, 00133 Rome, Italy}
	
	\author{M.\ Cencini}
	\affiliation{Istituto  dei  Sistemi  Complessi,  CNR,  via  dei  Taurini  19,  I-00185  Rome,  Italy}
	\affiliation{INFN ``Tor Vergata'' Via della Ricerca Scientifica 1, 00133 Roma, Italy}
	
	\author{G.\ Gallavotti}
	\affiliation{INFN, Sezione di Roma and Università ``La Sapienza'', Piazzale Aldo Moro 2, 00185 Roma, Italy \& Accademia dei Lincei,  Rome,  Italy}
	
	\author{V. Lucarini}%
	\affiliation{Department of Mathematics and Statistics, University of Reading, Reading, United Kingdom}
	\affiliation{Centre for the Mathematics of Planet Earth, University of Reading, Reading, United Kingdom}

	\date{\today}
	
	
	\begin{abstract}
		At the molecular level fluid motions are, by first principles, described by time reversible laws. On the other hand, the coarse grained macroscopic evolution is suitably described by the Navier-Stokes equations, which are inherently irreversible, due to the dissipation term. Here, a reversible version of three-dimensional Navier-Stokes is studied, by introducing a fluctuating viscosity constructed in such a way that enstrophy is conserved, along the lines of the paradigm of microcanonical versus canonical treatment in equilibrium statistical mechanics. 	Through systematic simulations we attack two important questions: (a) What are the conditions that must be satisfied in order to have a statistical equivalence between the two non-equilibrium ensembles? (b) What is the empirical distribution of the fluctuating viscosity observed by changing the Reynolds number and the number of modes used in the discretization of the evolution equation? The latter point is important also to establish regularity conditions for the reversible equations. We find that the probability to observe negative values of the fluctuating viscosity becomes very quickly extremely small when increasing the effective Reynolds number of the flow in the fully resolved hydro dynamical regime, at difference from what was observed previously.\footnote{Post-print version of the article published in Phys. Rev. E, 105(6), ~\href{https://doi.org/10.1103/PhysRevE.105.065110}{\underline{065110}}, (2022).}

	\end{abstract}
	
	\maketitle
	
	
	\def\SEC{Introduction}
	\section{\SEC}
	
	\label{sec1}
	
	The statistical balance between energy injection and dissipation is the key ingredient for the establishment of steady state conditions in non-equilibrium statistical mechanical systems. Such systems are driven out of equilibrium by the presence of an external forcing, while dissipation acts as a thermostat that removes the excess of energy \cite{Ga013b,Ga020b}, with entropy being produced in the process.
	
	In the case of fluid systems, described by the Navier-Stokes equation (NSE) \cite{Frisch1995,Pope2000}, dissipation is introduced in the form of a Laplacian operator acting on the velocity field times a positive constant: the viscosity. Such an operator preferentially damps the small scales of the flow.
	There are other ways to introduce dissipation, and two relevant examples are given next. For instance, in two-dimensional (2D) and geophysical flows, dissipation is often introduced via the Ekman friction, which is justified, e.g., by the effects of the bottom and top surfaces of the thin fluid layer which helps avoiding accumulation of energy at large scales \cite{Salmon1998,alexakis2018cascades}. Another approach, used customarily in numerical simulations to reach high intensity turbulent states in many applications, is to introduce hyperviscosity \cite{Frisch2008}, \ie, a positive power of the Laplacian, which confines dissipation to the small scales, acting, \textit{de facto}, like a sharp ultraviolet filter. It is also possible to introduce dissipation by combining more than one of the above  outlined approaches. The common characteristic among all the forms of dissipation described above is that they break the time reversal symmetry, which is instead preserved by the other terms of the NSE.  
	
	Reversible equations govern microscopic motion while irreversible equations describe, very often, macroscopic evolution of the same systems with equations derived via scaling of various parameters \cite{Cercignani1988,Zwanzig2001,OM953a,Bertini2002}. Thus, the question arises as to whether the macroscopic description of systems evolving irreversibly could also be described macroscopically by reversible equations.

	The equivalence of different ensembles in equilibrium statistical mechanics (see, e.g., \cite{Ru1999,Touchette2015}) describes, in the thermodynamic limit, the independence of macroscopic observables with respect to the chosen thermostat, which defines the underlying microscopic interactions between the system and the reservoir in contact with it. In non-equilibrium systems it is possible to appropriately modify the irreversible term(s) of the evolution equation in such a way that time reversibility is restored,  while, under suitable constraints, a macroscopic quantity is kept
	fixed \cite{EM990}; an early example for NSE is in
	\cite{SJ993} where many conditions are simultaneously imposed to constrain that the energy content obeys at every scale (above the Kolmogorov's) the $\frac{5}{3}$ law. The key question is whether, and under which conditions, the two ensembles (irreversible and reversible) are equivalent and describe the same physical problem.

	In the context of fluid systems, where the viscous term is the source of irreversibility, it is possible to introduce a velocity field-dependent fluctuating viscosity, such that the viscous term becomes formally reversible, while keeping fixed a macroscopic quantity (e.g.~the energy or enstrophy, etc.) as a constraint. The equivalence between the  irreversible and reversible fluid ensembles was first conjectured in \cite{Ga996b,Ga997b} in the limit of vanishing viscosity and fixed system size. Subsequent numerical attempts addressed this conjecture in a few simple systems such as in a version of the  2D NSE truncated to a few modes and with periodic boundary conditions \cite{GRS004}, and more recently in \cite{Ga020a,Ga020b}, in the Lorenz-96 model \cite{GL014} and a shell model of turbulence \cite{BCDGL018,DBBC018}. Recently, an interesting attempt to show equivalence between a reversible 2D NSE, where both energy and enstrophy are kept fixed, with irreversible 2D NSE is presented in \cite{SSMGS21}. First in \cite{Ga018} and later in \cite{Ga019c,Ga020b,Ga021a} a second  conjecture was laid to address the equivalence in the limit of infinite system size at fixed viscosity.

	Attempts to investigate the properties  of the reversible ensemble for the 3D NSE have only very recently been made \cite{SDNKT018,JC020}. In \cite{SDNKT018}, where the fluctuating viscosity is constructed in such a way as to keep the energy at a constant value, the authors gave insight of a possible second order phase transition between the two statistically steady regimes that 3D NSE can exhibit (see Sec.~\ref{sec:stat_regimes}), in the dual limit of infinite system size and vanishing viscosity, and gave a suggestion of the parametric space where the conjectures should be addressed. In \cite{JC020}, where enstrophy is kept constant, the authors employ high spatial resolutions at small viscosities and by comparing the expectation values of several different observables provide some evidences of the agreement between the generated ensembles of irreversible and reversible 3D NSE.

	The goal of this work is twofold. First, we wish  to clarify the content and study the domain of validity of the two conjectures for the 3D NSE, by thoroughly investigating the distributions of several observables at different scales, with attention to expectation values and standard deviations. Second, we wish to study the statistics of the fluctuating viscosity in the statistically steady regimes, which provides insights for the conditions of smoothness of the velocity fields.

	The paper is structured as follows. In Sec.~\ref{sec2} we present the necessary theory and state the conjectures we wish to test in our study. In Sec.~\ref{sec3} the results of the numerical simulations are provided. The results pertaining the two conjectures are separately addressed in Secs.~\ref{sec:Conj1_test} and \ref{sec:conj2_hydro}, respectively. In Sec.~\ref{sec4} we discuss the results on the statistics of the fluctuating viscosity for the reversible NSE. We summarize our findings and provide some perspectives in Sec.~\ref{sec:Conclusions}. We further supplement our work with a series of appendixes.

	\def\SEC{Viscosity and reversibility}
	\section{\SEC}
	\label{sec2}
	
	\subsection{Irreversible Navier-Stokes equation in 3D}
	
	Here we consider the classical case of an incompressible fluid enclosed in a three-dimensional container with periodic boundary conditions, described by the NSE. Assuming incompressibility amounts to removing internal energy from the energy content of the fluid.
	
	The NSE can be written as
	\begin{equation}
		\partial_t{\V u} =	\n\D\V u - { {\uu}}
		\cdot\BDpr\,\uu- \BDpr p+\V f, \Eq{e1.1a}
	\end{equation}
	where $\V u(\V x,t)$ is imposed to have zero divergence and zero spatial average, $p$ is the pressure field, $\n$ is the viscosity, and $\V f$ is a force field. The NSE is fundamentally difficult in 3D because it is not known whether solutions could be constructed with sufficient generality. For instance, if we suppose that $\ff$ acts only at large scales (\ie, confined to a finite number of modes) and the initial data have only a finite number of modes, then it is not known whether, in such a generality and no matter how small $\n>0$ is fixed, a smooth solution follows for all $t>0$. Note that this formulation of the problem is slightly weaker than the millennium problem B of the Clay Mathematics Institute \cite{Fe000}.
	
	On the other hand, equally hard problems arise in statistical mechanics systems; for instance, in 3D there is no existence theorem for an infinite system of hard spheres starting at an initial state in which the maximal speed and the minimal particles distance in any unit box are bounded away from $\infty$ and $0$. Notwithstanding, this has not been an obstacle for developing the statistical mechanical theory of phase transitions.

	The obvious way forward is to introduce extra parameters which regularize the equations, turning them into equations which admit solutions  and try to study only properties that can be shown to be independent of the regularization parameters. For instance, in the case of statistical mechanics the equations are typically regularized by confining the system to a finite box, say a cube of volume $V$.
	
	In the present case we consider the truncated NSE, \ie, the regularized version of Eq.~\equ{e1.1a} in Fourier space obtained by requiring that $\uu$ is periodic in the container and has only modes $\kk=(k_1,k_2,k_3)$ with  $ k \le N$,  and $k\defi||\kk||_2 =\sqrt{k_1^2+k_2^2+k_3^2}$. We focus on properties of the solutions which hold uniformly in the cut-off $N$, and the container is supposed to be of size $[0,2\p]^3$.  Therefore, one can introduce the complex scalars $u_{\b,\kk}=\lis u_{\b,-\kk}$, where the overline notation denotes the complex conjugate, and two unit vectors $\eE_\b(\kk)=-\eE_\b(-\kk)$ mutually orthogonal and orthogonal to $\kk$, 
	so that the velocity field can be written as
	\be
	\uu(\xx,t)=\sum_{\b=1,2; \kk}
	i\,u_{\b,\kk}(t) \,\eE_{\b}(\kk) e^{-i\kk\cdot\xx}, 
	\Eq{e1.2}
	\ee
	%
	where the sum is restricted to $k \le N$. 
	Furthermore, by defining the kernel $ T_{\pp,\qq,\kk}^{\b_1,\b_2,\b}
	=-(\V e_{\b_1}(\pp)\cdot \qq) (\V e_{\b_2}(\qq)\cdot \V
	e_{\b}(\kk)), $
	and making the time notation implicit unless strictly necessary, the NSE can be expressed as
	\begin{equation}
		\partial_t u_{\b,\kk}=\sum_{\b_1,\b_2\atop\kk=\pp+\qq}
		T_{\pp,\qq,\kk}^{\b_1,\b_2,\b}\,u_{\b_1,\pp}u_{\b_2,\qq} -\n k^2u_{\b,\kk}+ f_{\b,\kk},\label{e1.4} 
	\end{equation}
	%
	with $k, p, q \le N$.
	Due to the symmetry between $\qq,\kk$, the sum over $\b_1,\b_2$ keeping $\pp+\qq+\kk=\V0$ yields the identity
	\begin{equation}
		\sum_{\b_1,\b_2}
		T_{\pp,\qq,\kk}^{\b_1,\b_2,\b_2}\,u_{\b_1,\pp} u_{\b_2,\qq}
		u_{\b_2,\kk}=0,\Eq{e1.5}
	\end{equation}
	reflecting the conservation of both the total energy, $\int d \xx \, \uu^2$, and the total helicity, $\int d\xx \, \uu\cdot(\BDpr\wedge\uu)$ in the unforced and inviscid ($\n=0,\ff=\V0$) limits.

	\subsection{Reversible Navier-Stokes Equation in 3D}

	Historically, discussions on the origin of dissipative macroscopic properties, out of purely reversible Newtonian dynamics at molecular level, go back at least to Maxwell [see Eq.~(128) in \cite{Maxwell2011}]. Therefore, it is natural to expect that fluid motion can also be described by microscopic reversible equations. One can thus inquire whether even at the macroscopic level, although molecular motion is no longer explicitly playing a role, the same phenomena can be described by macroscopic reversible equations.

	Here we shall study stationary states of the truncated NSE. The basic idea is that viscosity controls the regularity of the flow by forbidding uncontrolled growth of energy, and dissipates the input work by the forcing: $\int \ff\cdot\uu \,d\xx$, proportionally to the product of viscosity times {\it enstrophy}   
	\begin{equation}
		\n \DD(\uu) = \n \sum_{\kk,\b} k^2|u_{\b,\kk}|^2,
		\label{eq:nuD}
	\end{equation}
	and where at steady state there is a statistical balance between input and output of energy.
	One can envisage that by replacing the viscous force $\n\D\uu$ with a reversible force that keeps the enstrophy constant many statistical properties of stationary states will remain correctly described, independently of the cut-off $N$ \cite{Ga006c}.
	
	The analogy with statistical mechanics is helpful; there, the cut-off is the volume of the container, and the equilibrium states can be described by different microscopic dynamics, like the energy conserving microcanonical distribution or the isokinetic evolution \cite{EM990}. Both choices of states and evolution laws give the right statistics for {\it local observables}, \ie, observables depending on the configurations of particles located in a volume small compared to the total one, \ie, to the cut-off $V$. Most importantly, the statistical properties of such observables depend on $V$ less and less as $V\to\infty$.
	
	In the case of NS fluids subject to large scale forcing, locality is defined with respect to the Fourier space. The analogs of local observables are functions of the velocity field $\uu$ that only depend on components $\uu_\kk$ with $k$ small compared to the cut-off $N$. More specifically, $O(\uu)$ is a local observable,  when it depends only on a finite number of modes with $k <K$ for some $K < N$.  This leads to the proposal that replacing the standard viscous force with a dissipative reversible term that keeps the enstrophy constant will lead to stationary states completely equivalent to the original dynamics up to an arbitrary $K$, as far as the statistical properties of local observables are concerned, provided $N$ is large (ideally $N \to \infty$, in the same sense in which equilibrium thermodynamics becomes ensemble independent as the volume tends to infinity).  Here $K$ can in principle depend on the viscosity and it is important to understand its scaling properties in the $\nu \to 0$ limit (see below).
	
	Specifically, we consider a different evolution equation where 
	(i) in Eq.~\equ{e1.4} the external force $\ff$ is fixed, with $||\ff||_2=\text{const}\in\mathbb{R}$, and
	acts on ``large scale'', \ie, it has finitely many modes: $\ff_\kk=0$
	if $k\ge k_f$ for some $k_f$; (ii) the viscous force $-\nu k^2 \uu_\kk$ is replaced by $-\a(\uu)k^2\uu_\kk$, with $\a(\uu)$ defined such that the enstrophy $$D=\DD(\uu)$$ is a constant of motion. Note that a similar choice was followed by the authors in \cite{JC020}, while in \cite{SDNKT018} the kinetic energy was instead kept fixed. Such assumptions lead to the equations
	\begin{equation}
		\partial_t u_{\b,\kk}= \sum_{\b_1,\b_2\atop\kk=\pp+\qq}\kern-3mm
		T_{\pp,\qq,\kk}^{\b_1,\b_2,\b}\ u_{\b_1,\pp}\,u_{\b_2,\qq}
		-\a(\uu)k^2u_{\b,\kk}+ f_{\b,\kk},
		\Eq{e2.1}
	\end{equation}
	where
	\be 
	\a(\uu)=\frac{\L(\uu)+\sum_{\b,\kk}k^2 f_{\b,\kk}\lis u_{\b,\kk}}
	{\sum_{\b,\kk} \kk^4|u_{\b,\kk}|^2} = \frac{\Lambda(\uu) + W(\uu)}{\Gamma(\uu)},	
	\Eq{e2.2}\ee
	with
	\begin{equation}
		\L(\uu)=\sum_{\b_1,\b_2\atop\kk=\pp+\qq}
		(T_{\pp,\qq,\kk}^{\b_1,\b_2,\b_2}\, k^2\,
		u_{\b_1,\pp}u_{\b_2,\qq}
		\lis u_{\b_2,\kk})
		\Eq{e2.3}
	\end{equation}
	obtained by multiplying both sides of Eq.~\equ{e2.1} by $k^2 \lis u_{\b,\kk}$, then summing over $\kk$, and finally imposing that the right-hand side equals 0.

	We call Eq.~\equ{e1.4} irreversible NSE (abbreviated as $\I$) and Eq.~\equ{e2.1} reversible NSE (abbreviated as $\R$). The name reversible refers to the property that if $S_t\uu=\uu(t)$ is a solution of $\R$, with initial data $\uu_0$, and if ${\cal P}\uu=-\uu$, then ${\cal P}S_t\uu=S_{-t}{\cal P}\uu$ is also a solution, as a consequence of $\a({\cal P}\uu)=-\a(\uu)$. Instead, such an identity does not hold for the solutions of $\I$.
	
	\subsection{Conjectures}\label{sec:conjectures}
	
	Let us introduce the collection $\EE^{\I,N}$ of the stationary distributions $\m^{\I,N}_\nu$ for the $\I$\  evolutions, with cut-off $N$; for a given choice of $\ff$ the distributions are parametrized by the Reynolds number $\re\propto\n^{-1}$. In principle, several distributions may correspond to the same $\re$ although it is plausible that for $\re$ large enough there is only one stationary distribution. Likewise, let $\EE^{\R,N}$ be the collections of the stationary distributions $\m^{\R,N}_D$ for the $\R$ evolutions, with the same cut-off $N$. These are parametrized by the value $D$ of the (constant) enstrophy. Again, in general, several distributions may correspond to the same $D$; similarly to the irreversible case, it is plausible that for generic initial data and $D$ large,  $\mu_D^{R,N}$ is unique.
	
	It is natural to associate to each distribution in $\EE^{\I,N}$
	and $\EE^{\R,N}$ the average enstrophy $\media{\DD(\uu)}_\n^{\I,N}$
	and, respectively, the enstrophy $D$, as well as the
	work per unit time dissipated:
	\be \mathcal{W}^{\I,N}_\n= \int
	d\xx\, \langle\ff\cdot\uu\rangle^{\I,N}_\nu, \quad
	\mathcal{W}^{\R,N}_D=\int d\xx\langle\ff\cdot\uu\rangle^{\R,N}_D
	\Eq{e2.4a}\ee
	where $\media{O}^{\I,N}_\nu$, $\media{O}^{\R,N}_D$ denote the
	averages of an observable $O$ over the distributions
	$\m^{\I,N}_\nu$ and $\m^{\R,N}_D$. Parameters $\n,D$ at given
	$N$, will be said to be in correspondence if
	\begin{equation}
		\media{\DD(\uu)}_\n^{\I,N}=D.
		\label{e2.4}
	\end{equation} 
Equation \eqref{e2.4} defines an implicit relationship between $D$ in $\R$ and $\nu$ in $\I$.
	
	In this work, we perform a few preliminary checks, based on a numerical analysis of the truncated NSE, of the following	two conjectures on the distributions of local observables, originally presented in \cite{Ga997b,GRS004,Ga013b}, respectively.\\ 
	
	\0{\bf Conjecture 1:}  If the parameters $\n,D$ are in the correspondence defined in Eq.~\eqref{e2.4}, then for all observables $O(\uu)$ one has
	\be
	\frac{ \lim_{\nu\to 0} \media{O}^{\R,N}_D}{\lim_{\n\to0} \media{O}^{\I,N}_\nu} = 1,
	\Eq{e2.5}\ee
	for all $N$.  \*
	\0{\it Remark.}  In the case where  the mean value of the observable is zero or infinity, Eq.~\eqref{e2.5}  is intended to mean that the same value is obtained in both ensembles. In such cases where there are several invariant distributions with the same $\nu$ or same $D$, Eq.~\eqref{e2.5} has to be interpreted as implying that a correspondence can be set between a pair of distributions for each collection.  conjecture 1 was proposed and tested for different systems in a limited number of cases (\eg, \cite{Ga997b,GL014,BCDGL018}).
	
 Let us introduce the {\it Kolmogorov scale} $k_\nu = \left(\frac{\varepsilon}{\nu^3}\right)^{1/4}$, with $\varepsilon=\n \langle\DD(\uu)\rangle$ being the energy dissipation rate, defined as the typical length where inertial and {\it irreversible} dissipative effects balance. We can then state  a different  conjecture in the limit of $N \to \infty$.
	
	\* \0{\bf Conjecture 2 -- Equivalence
		Hypothesis:}  Let $O$ be a local observable, \ie, a function of $\uu$
	depending only on a finite number of modes with $k
	<K$. Then if $\m^{\I,N}_\nu$ and $\m^{\R,N}_D$ satisfy
	Eq.~\equ{e2.4} one has
	\be \frac{ \lim_{N\to\infty} \media{O}^{\R,N}_D}{\lim_{N\to\infty}
		\media{O}^{\I,N}_\nu } = 1\Eq{e2.6}\ee
	for all $\n$ and $K<c_\n k_\n<N$ with $0<c_\n\to c_0\le\infty$ as $\n \to 0$.  It is important to notice that  in the case $K \to \infty $ when $\nu \to 0$ we have that the equivalence holds on a formally infinite dimensional manifold.

	Remarks: 
	\begin{enumerate}
		\item Conjecture 1 relies on the chaoticity of the evolution at
		small $\nu$ and fixed $N$. Instead, conjecture 2 concerns only
		the limit $N\to\infty$ and relies on the chaoticity of the
		microscopic motions leading to the NSE. Conjecture 2 is
		formulated for all $\n$, including the case where the
		asymptotic motion consists of periodic attracting sets. It just
		provides a quantitative version of conjecture 1: given a local
		observable living on scale $K$ the question of how small should
		$\n$ be to achieve equivalence receives the quantitative answer
		that, in the limit $N\to\infty$, $\n$ has to be such that
		$K<c_0 k_\n$ which becomes eventually true, as lower and lower values of $\nu$ are considered, because $k_\n$
		diverges as $\n\to0$.
		
		\item Conjecture 2 could be extended to more general macroscopic
		equations derived via scaling limits from microscopic
		reversible evolution.
		
		\item In cases with more than one distribution for a given $\nu$
		or $D$ the interpretation should be that extra labels should be
		added to distinguish the various distributions and follow the
		remark in conjecture 1. Conjecture 2 first appeared in
		\cite{Ga018} and was formally proposed in \cite{Ga019c}.

		\item In \cite{Ga020b,Ga021a} a stronger conjecture is directly formulated with $c_\nu=\infty$ for all $\nu$.  \label{item4conj2}
	\end{enumerate}

	The two conjectures differ in the order of consideration of
	the two limits, $\nu \to 0$ and $N \to \infty$. In particular, it
	is important to stress that the so-called fully developed
	turbulent limit is achieved in nature by $N \to \infty$ first and
	then $\nu \to 0$, \ie, it is captured by conjecture 2. On the
	other hand, the limit $\nu \to 0$ for fixed $N$ leads to a
	quasi-thermalization (in the sense of \cite{AB020}) of the high
	wave numbers range cut-offed by the maximum $N$. An additional
	feature of this regime is that the dimension of the attractor
	approaches the number of degrees of freedom of the system
	\cite{GL014}.

	\def\SEC{Numerical Simulations}
	\section{\SEC}
	\label{sec3}
	\subsection{Setup}
	
	We have performed direct numerical simulations of both $\I$ and $\R$ equations for incompressible fluid in a triply periodic domain of size $L= 2\pi$. We used a dealiased (following the $\frac{2}{3}$ rule \cite{Orszag1971}) parallel 3D pseudospectral code, based on the P3DFFT implementation \cite{P3DFFT}, on cubic grids of size $N_0 = 64$, $128$, $256$, and $512$ collocation points in each direction, effectively corresponding, via $N=N_0/3$,  to $N=21$, $42$, $85$, and $170$ in Fourier space, respectively. $N$ is the same as the cut-off scale $k_{max}$ and will be used interchangeably. The time integration has been implemented with a second-order Adams-Bashforth scheme and a small timestep equal to $\Delta t=10^{-4}$ was chosen for all the simulations; tests with different time steps have been done confirming that the one chosen was sufficient to ensure the robustness of the presented results. The grid size is $\Delta x=2\pi/N_0$. The zero mode is not forced, ensuring that the mean velocity field remains zero at all times, and the external forcing is deterministic and only acts on large scales, specifically in the $k_f\in[1,\sqrt{2}]$ domain. The phases were chosen as quenched random variables and kept equal in all simulations with $\sum_\kk k^2 ||\V f||_2^2=1$. We refer to Table \ref{tab:parameters} in Appendix \ref{appendixTable} for a summary of the numerical simulations we performed, which are labeled as  R\#, where R stands for run in this context, followed by the number of the run, and a symbol, indicating the statistical regime each run belongs to; see discussion in Sec.~\ref{sec:stat_regimes}.

	We examine conjectures 1 and 2 by testing a set of viscosities: $\nu = 5\times 10^{-2}, 10^{-2},10^{-3},10^{-4},10^{-5}$ for different values of $N$. We proceed as follows. First we run a reasonably long $\I$ simulation at fixed $\nu$, $N$ and measure the average enstrophy $\langle D\rangle^{\I,N}_\nu$ in the statistically steady state. Next, we start the $\R$ run from a steady $\I$ state for which the instantaneous enstrophy is $D=\langle D\rangle^{\I,N}_\nu$. This step reduces the transient time required to reach the attracting set, but it is not strictly necessary, and in principle one can start from any initial condition with the appropriate enstrophy. We run the $\R$ for the same duration as the $\I$, and at each timestep we make tiny corrections by rescaling the velocity field as $\uu \rightarrow \uu \sqrt{\frac{\langle{D}\rangle^{\I,N}_\nu}{D(t)}}$. Although, given the very small timestep we use, the deviations from condition \eqref{e2.4} were tested to be small, the latter correction ensures that the total enstrophy stays constant within machine precision.

	\subsection{Statistically steady regimes}
	\label{sec:stat_regimes}
	
	Two distinct statistically steady regimes can be identified for the $\I$ and $\R$ systems, depending on the cut-off $N$ and Re (or, equivalently, on $k_\nu$). One is what we call the \textit{hydrodynamic} regime, achievable when $N$ is large enough for any $\nu$, such that $k_\nu \lesssim N$. It is characterized by a well developed inertial range (for small $\nu$) with a $E(k){\sim}k^{-5/3}$ scaling for $k_f \ll k \ll k_\nu$ and a well resolved exponential or superexponential decay for large wave numbers, where 
	\begin{equation}
		E(k)=\sum_{\beta,k-\frac{1}{2}<k<k+\frac{1}{2}}\langle |u_{\beta,\kk}|^2\rangle
		\label{eq:E_k}
	\end{equation}
	are the averaged spectral properties, or simply the energy spectrum. A scaling close to $E(k){\sim}k^{-5/3}$ is  observed in nature because of the regularizing properties of viscosity at small scales \cite{Frisch1995,Pope2000,alexakis2018cascades}. The other, -\textit{quasithermalized}- regime \cite{AB020}, features a $E(k){\sim}k^{2}$ scaling for large values of $k$ \cite{AB020} and it is obtained at sufficiently small $\nu$ for any fixed $N$, such that $k_\nu \gg N$. We call this regime quasithermalized because the average energy content of the individual modes depends only weakly on $k$, given the geometrical degeneracy of the energy shells.  A crossover region can be located between the two in the ($N,\nu$) parameter space \cite{SDNKT018}.

	\begin{figure}[!h]
		\centering
		\includegraphics[width=.9\linewidth]{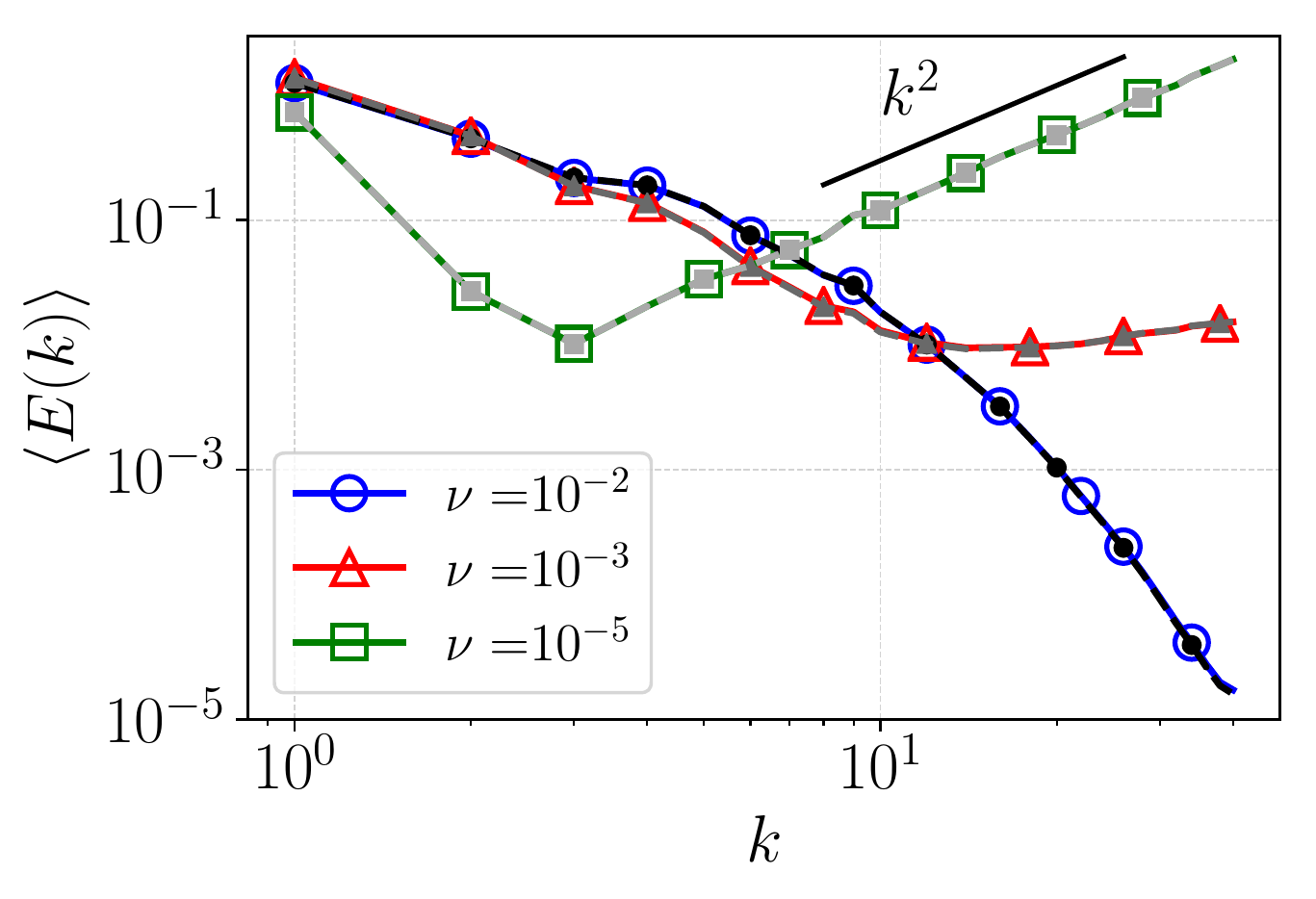}
		\caption{Energy spectrum of $\I$ (grayscale filled symbols) and $\R$ (colored open symbols) for $N_0=128$ and at changing $\nu$ (for $\I$) or $D$ (for $\R$). We observe a well developed hydrodynamic regime for $\nu=10^{-2}$ (R$7\bigcirc$),  a  crossover regime for  $\nu=10^{-3}$ ( R$8\triangle$) and a quasithermalized regime for $\nu=10^{-5}$ (R$10\square$); see Table \ref{tab:parameters} in Appendix \ref{appendixTable} for details on the runs. }
		\label{fig:energy_spectrum} 
	\end{figure}
	
	In Fig.~\ref{fig:energy_spectrum} we present a first comparison between $\I$ and $\R$ by looking at  $E(k)$ at changing $\nu$ and for  fixed  spatial resolution, $N_0=128$. Here,  we have omitted in $\langle \cdot \rangle$ the label $\I$ and $\R$ for simplicity and the same will be done in what follows whenever it does not lead to ambiguities. All regimes can be accessed as the viscosity is varied, and in terms of the average energy spectrum, the $\I$ and $\R$ agree remarkably well. This will be further checked in the following with respect to the two conjectures. We anticipate that deviations will appear at scales $k$ beyond the Kolmogorov scale $k_\nu$, which will be further explored later. Next we will deal with the two regimes separately. 
	
	\subsection{Mean properties of $\alpha$}\label{sec:mean_alpha}
\begin{figure}[!h]
		\centering
		\includegraphics[width=.9\linewidth]{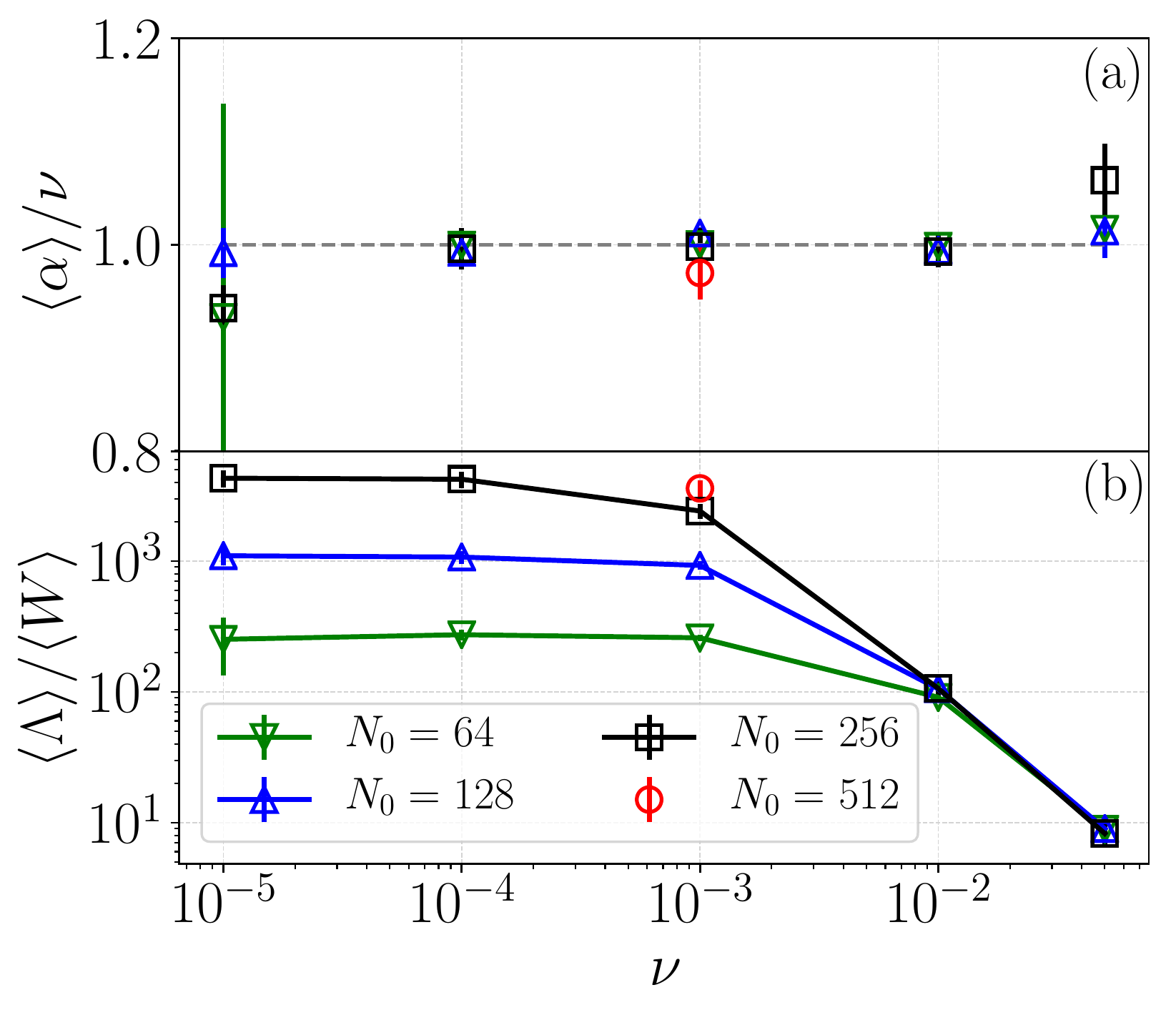}
		\caption{Mean properties of $\alpha$. (a) The ratio $\langle \alpha(\uu) \rangle^{\R,N}_D / \nu = 1$ is a consistency check for equivalence conditions. (b) The ratio $\media\L/\media{W}$ from Eq.~\eqref{e2.2} as a function of $\nu$ for all $N_0$ considered. The data are for $\R$. }
		\label{fig:mean_alpha_R} 
	\end{figure}

	A rigorous (yet non trivial, see below) consequence of both conjectures is the relation
	\begin{equation}
	\lim_{N\to\infty}	\langle \alpha(\uu) \rangle^{\R,N}_D = \nu,
		\label{eq:fl_visc}
	\end{equation}
	which holds when Eq.~\eqref{e2.4} holds. In Fig.~\ref{fig:mean_alpha_R}(a) we validate  Eq.~\eqref{eq:fl_visc}. The errors are calculated by $\delta \mathcal{O} = \sigma(\mathcal{O}) \sqrt{1+2\tau_{\mathcal{O}}}/\sqrt{j},$ where $\sigma(\mathcal{O})$ is the standard deviation, $\tau_{\mathcal{O}}$ the non-dimensional autocorrelation time, and $j$ the ensemble size.  
	
	Figure \ref{fig:mean_alpha_R}(a) provides a consistency check, which looks at first not entirely obvious because $\a$ is not a local observable;  see the definition given in Eq.~\eqref{e2.2}. Yet, if the conjectures and condition \eqref{e2.4} hold, it must be true that the ensemble average of $\a$ is equal to the viscosity. This follows from the balance between energy input and  output in steady state conditions that hold both for $\I$ and $\R$:
	\be \eqalign{  \lim_{N\to\infty}\n\, \media{\DD(\uu)}^{\I,N}_\n &=  \lim_{N\to\infty} \media{\ff\cdot\uu},
		\qquad \,\, \I\cr
		\lim_{N\to\infty} D\,\media{\a(\uu)}^{\R,N}_D &= \lim_{N\to\infty}\media{\ff\cdot\uu},\qquad \R.\cr}
	\Eq{e4.2}\ee
	Under the equivalence condition $\media{\DD(\uu)}^{\I,N}_\n=D$, the value $\lim_{N\to\infty}\media{\ff\cdot\uu}$ must be the same in $\I$ and $\R$, because $\ff\cdot\uu$ is a local observable. Hence, the relation $\media{\DD(\uu)}^{\I,N}_\n=D$ imposes the equality of the averages $\lim_{N\to\infty}\media{\ff\cdot\uu}$ for the $\I$ and $\R$ systems, which, in turn, implies the {\it non trivial} relation  (\ref{eq:fl_visc}). We remark that, as $\alpha(\uu)$ can be measured also in $\I$, we can also check its statistics. We observe that $\media{\a(\uu)}^{\I,N}_\nu \approx \nu$ in all statistical regimes, which is nontrivial because it is not a consequence of the equivalence conjecture because $\a$ is not a local observable. Although the distributions of $\alpha(\uu)$ in both $\R$ and $\I$ are in all cases peaked around $\nu$, as well as the mean values of both $\media{\a(\uu)}^{\R,N}_D$ and $\media{\a(\uu)}^{\I,N}_\nu$ are  approximately equal to $\nu$, their tails are different in the hydrodynamic regime, but become almost identical in the quasithermalized regime (see Sec.~\ref{sec:num_res_alpha_I}).

	Since $\a$ is a sum of the forcing contribution $W(\uu)/\G(\uu)$ and of the internal nonlinear contribution $\L(\uu)/\G(\uu)$ \eqref{e2.3}, it is interesting to remark that, on average, the non linear term dominates the forcing one as shown in Fig.~\ref{fig:mean_alpha_R}(b), where the ratio $\langle \Lambda \rangle / \langle W \rangle $ versus $\nu$ is plotted. The ratio increases linearly as $\nu$ decreases in the hydrodynamic regime, before it approaches a (large) constant depending on $N$ in the quasithermalized regime. Thus, on average the internal nonlinear exchanges dominate over the forcing for the $\R$ dynamics, introducing a non-local (in scale) coupling among all modes and making the validity of conjecture 2 even less trivial.

	\subsection{Test of conjecture 1:  Quasithermalized regime}\label{sec:Conj1_test}
	In this section we address the validity of conjecture 1 for 3D NSE. We remark that it has been confirmed in simpler dynamical systems in \cite{GRS004,GL014,BCDGL018,DBBC018}. conjecture 1 applies when we consider decreasing values of $\nu$ at fixed $N$, which leads to the quasithermalized regime. In terms of observables, we consider the single mode kinetic energy, $U(\kk)= ||\uu(k_1,k_2,k_3)||_2^2$ and the energy spectrum, $E(k)$ \eqref{eq:E_k} which will be presented separately.
	
	\subsubsection{Single mode energy}
\begin{figure}[!t]
		\centering
		\includegraphics[width=1.0\linewidth]{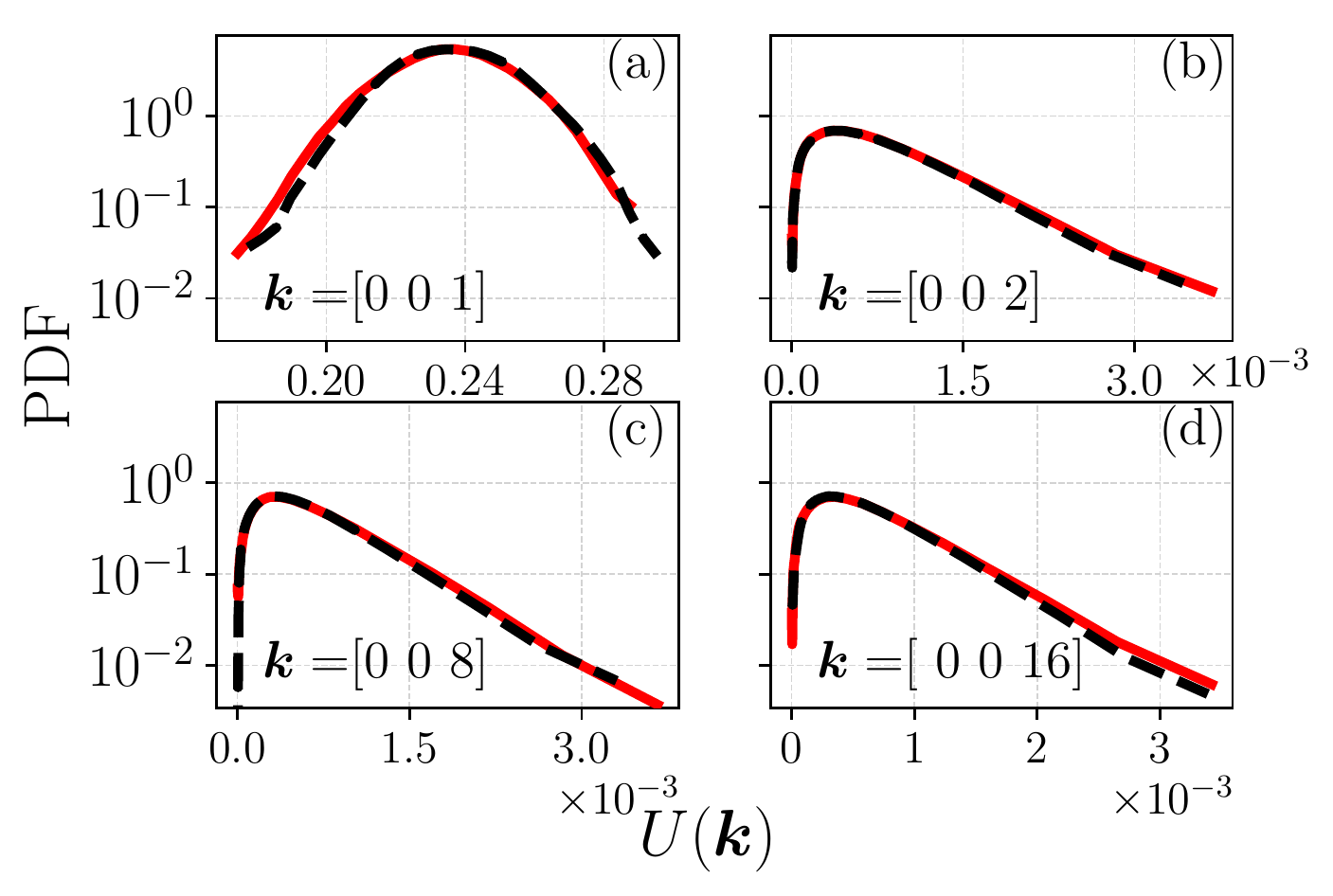}
		\caption{Probability distribution function of single mode kinetic energy $U(\kk)$ for chosen modes $\kk= (0,0,m)$, $m=1,2,8,16$, showing the equivalence between $\R$ (red straight lines) and $\I$ (black dashed lines). Here $\nu=10^{-4}$ and $N_0=64$ (R$4\square$), belonging to the quasithermalized regime. }
		\label{fig:modes_equiv} 
	\end{figure}

	We consider the probability distribution function (PDF) of the energy $U(\kk)$ of a certain mode $\kk$. In Fig.~\ref{fig:modes_equiv} we test this for a chosen type of modes, $\kk=(0,0,m)$, with $m=1, \,2, \,8, \, 16$, and we see a very good agreement between the two ensembles, where the output of $\R$ is shown in red straight lines, while black dashed lines are used for the $\I$ simulations. In our analysis we have considered several different types of modes, i.e.,permutations of $(0,0,m)$, $(1,0,m)$ and $(0,p,m)$, with $m=1,\ldots, N$ and $p=$ 2, or 7, or 9 (the latter integers were randomly chosen). The PDFs approximately follow a $\chi^2$ distribution. 
	
	\begin{figure}[!h]
		\centering	
		\includegraphics[width=1.\linewidth]{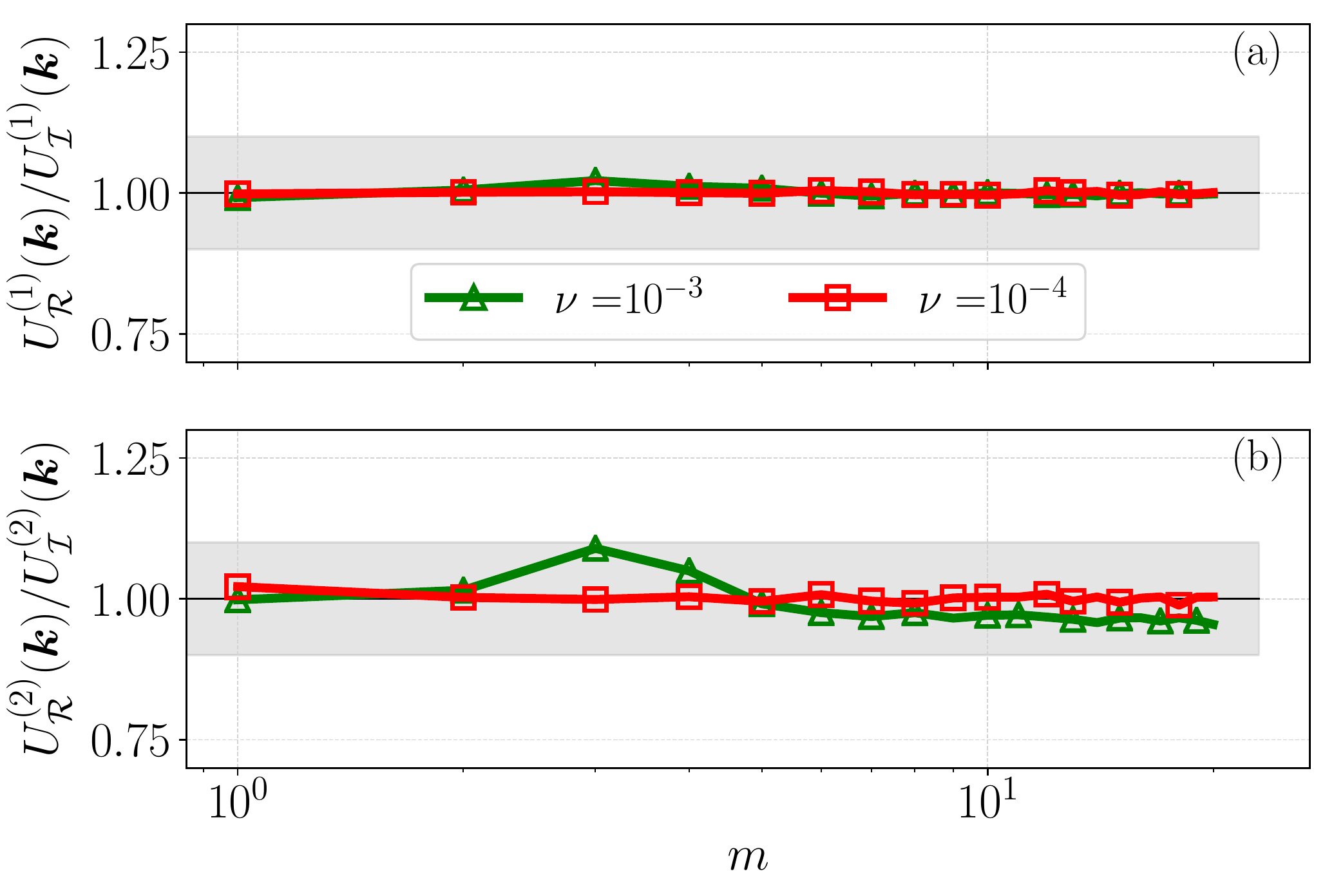}	
		\caption{Test of conjecture 1 for single modes kinetic energy $U(\kk)$. (a) Averaged $\R/\I$ ratio of the mean of kinetic energy of considered modes, \ie, permutations of $(0,0,m)$, $(1,0,m)$ and $(0,p,m)$, with $m=1,\ldots, N$ and $p=$ 2, or 7, or 9. (b) Same for standard deviation ratio. The gray band indicates a 10\% deviation from 1. Here $N_0=64$, and results from R$3\triangle$, R$4\square$ are shown. }
		\label{fig:modes_equiv_ratios_N_64} 
	\end{figure}
	
	To further quantify the comparison between $\I$ and $\R$, we study the statistical properties of the ensembles of $U(\kk)$ by computing the mean $U^{(1)}(\kk) \equiv \langle U(\kk) \rangle$, and the standard deviation $U^{(2)}(\kk) \equiv \sqrt{ \langle U^2(\kk) \rangle -  \langle U(\kk) \rangle^2 }$  of the time series for each measured mode $\kk$. An indication of equivalence is when the ratios of such quantities are close to 1. In Fig.~\ref{fig:modes_equiv_ratios_N_64} we present the ratios of the mean $U_{\R}^{(1)}(\kk) / U_{\I}^{(1)}(\kk)$ (a) and the ratios of the standard deviation $U_{\R}^{(2)}(\kk) / U_{\I}^{(2)}(\kk)$ (b). To increase statistical accuracy, we further average the resulting ratios of the different types of modes considered here at each $m=1,\ldots, N$, corresponding to the same shells $k$. As shown in Fig.~\ref{fig:modes_equiv_ratios_N_64}, by fixing $N_0=64$ while decreasing $\nu$, conjecture 1 holds well in the case of single mode kinetic energy, which is a prototypical {\it local} observable, for the run (R$4\square$). Note that the run (R$3\triangle$) belongs to the crossover regime, hence outside the domain of validity of conjecture 1, leading to some small deviations from 1 in Fig.~\ref{fig:modes_equiv_ratios_N_64}(b). As a confidence interval, we introduce on qualitative grounds a 10\% deviation from 1, displayed as a gray band.

	\subsubsection{Energy spectrum}

\begin{figure}[h]
		\centering
		\includegraphics[width=1.\linewidth]{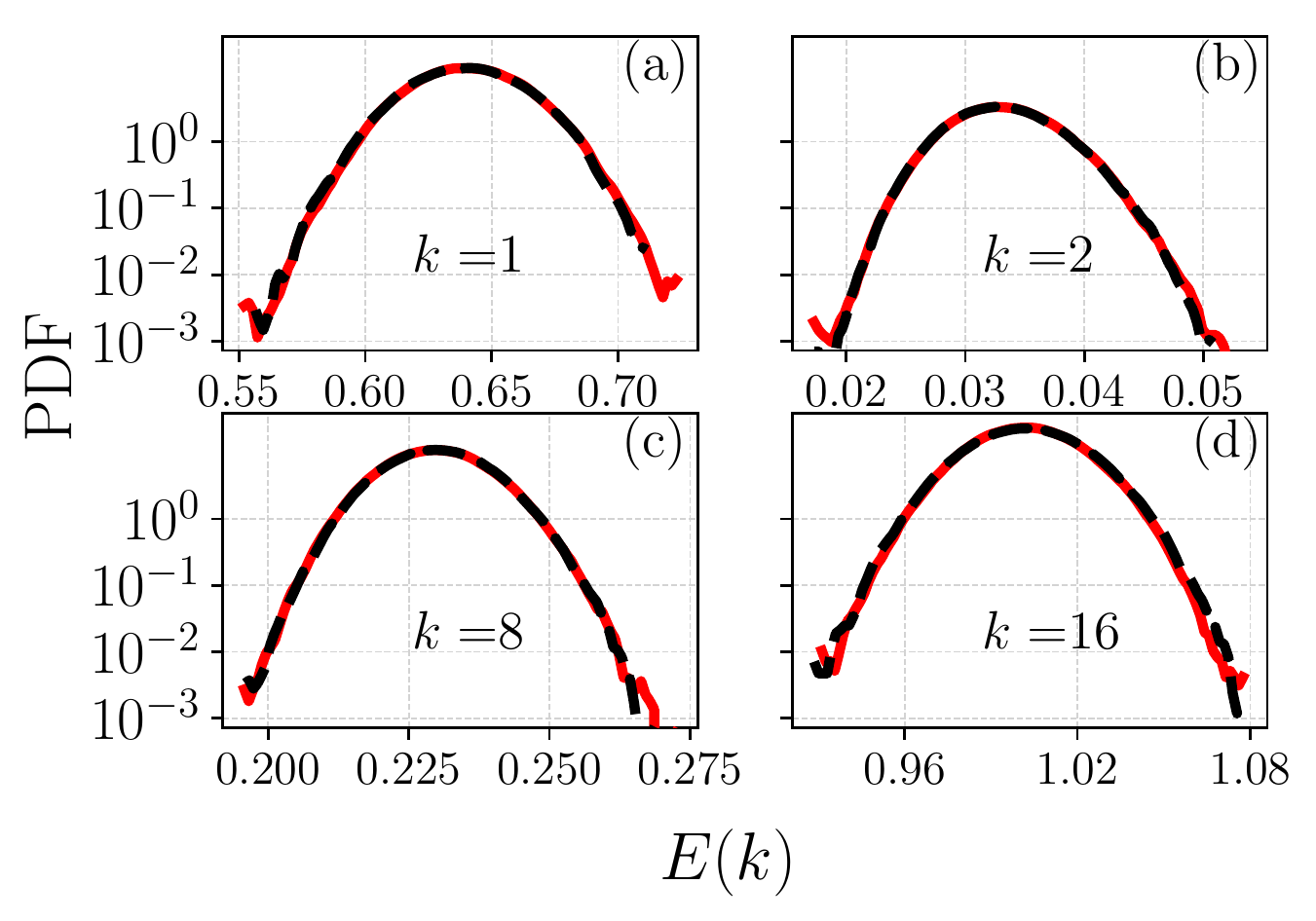}
		\caption{Probability distribution function of energy spectra for $k= 1, \,2, \,8, \,16$ showing the equivalence between $\R$ (red straight lines) and $\I$ (black dashed lines). Here $\nu=10^{-4}$ and $N_0=64$ (R$4\square$) corresponding to the quasithermalized regime.  }
		\label{fig:shells_equiv_qeq} 
	\end{figure}

	Subsequently, we study the energy spectrum, which has been briefly presented in Fig.~\ref{fig:energy_spectrum}. In Fig.~\ref{fig:shells_equiv_qeq} we show the PDF of kinetic energy for different shells $k= 1, \,2, \,8, \,16$ for the case $N_0=64$ and $\nu=10^{-4}$ (R$4\square$). We notice that the statistics of $\R$  and $\I$ are almost identical and are approximately Gaussian as previously observed in \cite{Brun2001}; {deviation from Gaussianity are nonetheless present for $k=2$.}
	
	\begin{figure}[h]
		\centering	
		\includegraphics[width=1.\linewidth]{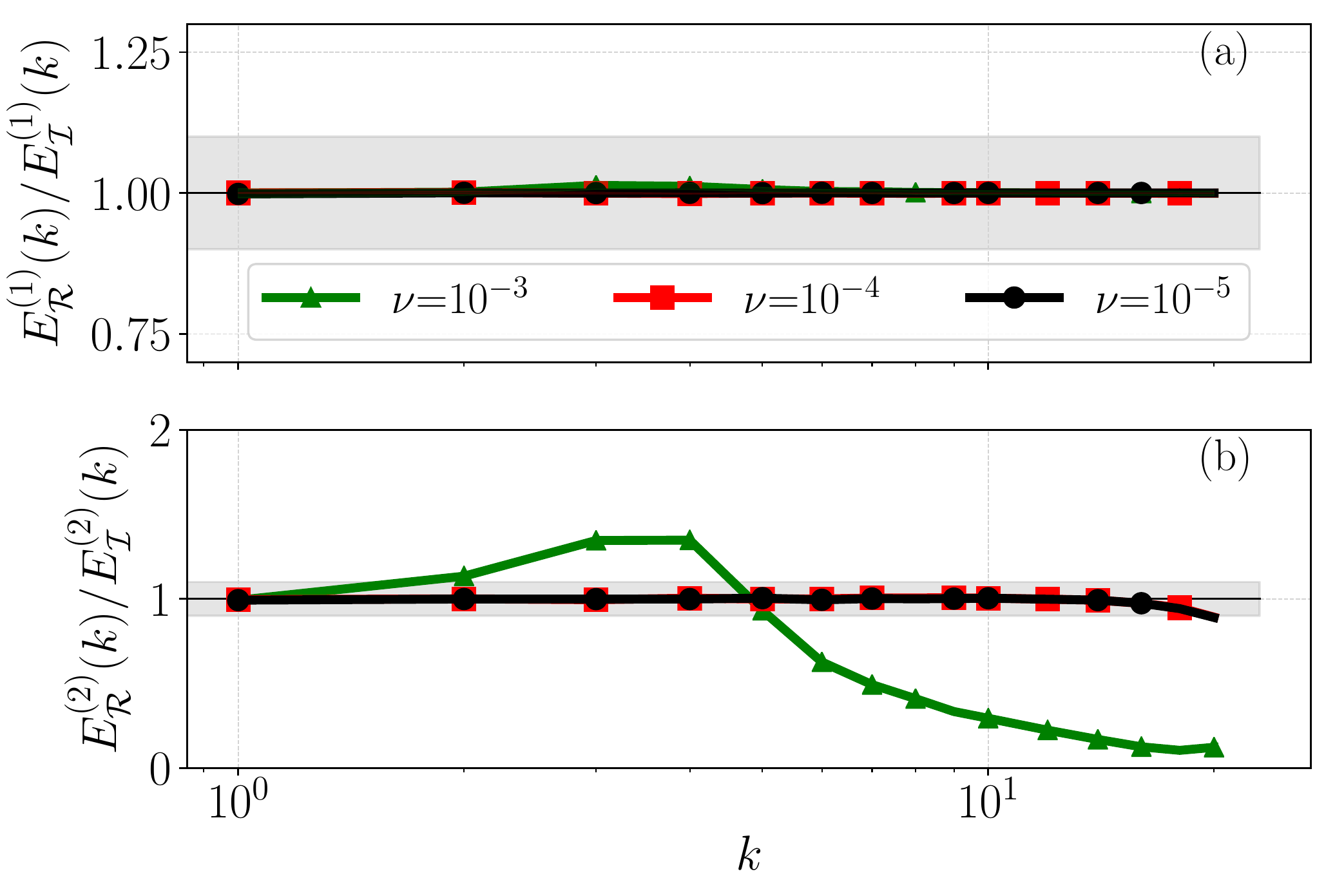}	
		\caption{Test of conjecture 1 for the energy spectrum. (a) $\R/\I$ ratio of mean energy spectrum at shell $k$ for $k=1,\ldots,N$. (b) Same for standard deviation ratio. The gray band indicates a 10\% deviation from 1. Here $N_0=64$, showing R$3\triangle$ ($\nu=10^{-3}$ - green triangles),  R$4\square$ ($\nu=10^{-4}$, red squares) and R$5\square$ ($\nu=10^{-5}$, black circles). Note that we are here also inspecting the crossover regime (green triangles), hence outside the validity of conjecture 1 [as shown in (b)]. Nonetheless, good agreement is still found in (a). }
		\label{fig:shells_equiv_ratios_conj1_qeq} 
	\end{figure}

	In Fig.~\ref{fig:shells_equiv_ratios_conj1_qeq} we summarize the equivalence between the two ensembles by plotting the same of Fig. \ref{fig:modes_equiv_ratios_N_64} but for the spectral properties at changing $k$ and at fixed $N_0=64$. Accordingly, we define the mean $E^{(1)}(k) \equiv \langle E(k) \rangle$, and the standard deviation $E^{(2)}(k) \equiv \sqrt{ \langle E^2(k) \rangle -  \langle E(k) \rangle^2 }$  of the time series for each shell $k$. As one can see, conjecture 1 is well verified at decreasing $\nu$ and for any  $k$. Notice in Fig.~\ref{fig:shells_equiv_ratios_conj1_qeq}(b) that the standard deviation ratios for $\nu=10^{-3}$ (green line-points) are different from 1 except for very small values of $k$. This is to be expected, as this case (R$3\triangle$) belongs to the crossover regime, hence outside the domain of validity of conjecture 1. Interestingly, for the same parameters, $\nu=10^{-3}$, $N_0=64$, the single mode standard deviation ratios are close to 1; see the green line in Fig.~\ref{fig:modes_equiv_ratios_N_64}(b). This reflects the higher complexity of the  energy spectrum, resulting from the presence of correlations among Fourier modes. All the above confirm the validity of conjecture 1.

	\subsection{Test of conjecture 2: Hydrodynamic regime} \label{sec:conj2_hydro}
	
	We now present our numerical tests of conjecture 2, which applies at fixed $\nu$ when $N$ is large, corresponding to the hydrodynamic regime.  We follow a systematic approach by first keeping the value of $\nu$ fixed, and then progressively increase the value of $N$. We then repeat this protocol for different values of $\nu$. Again, we consider the single mode kinetic energy and the energy spectrum.

	\subsubsection{Single mode energy}
	
	\begin{figure}[!h]
		\centering
		\includegraphics[width=1.\linewidth]{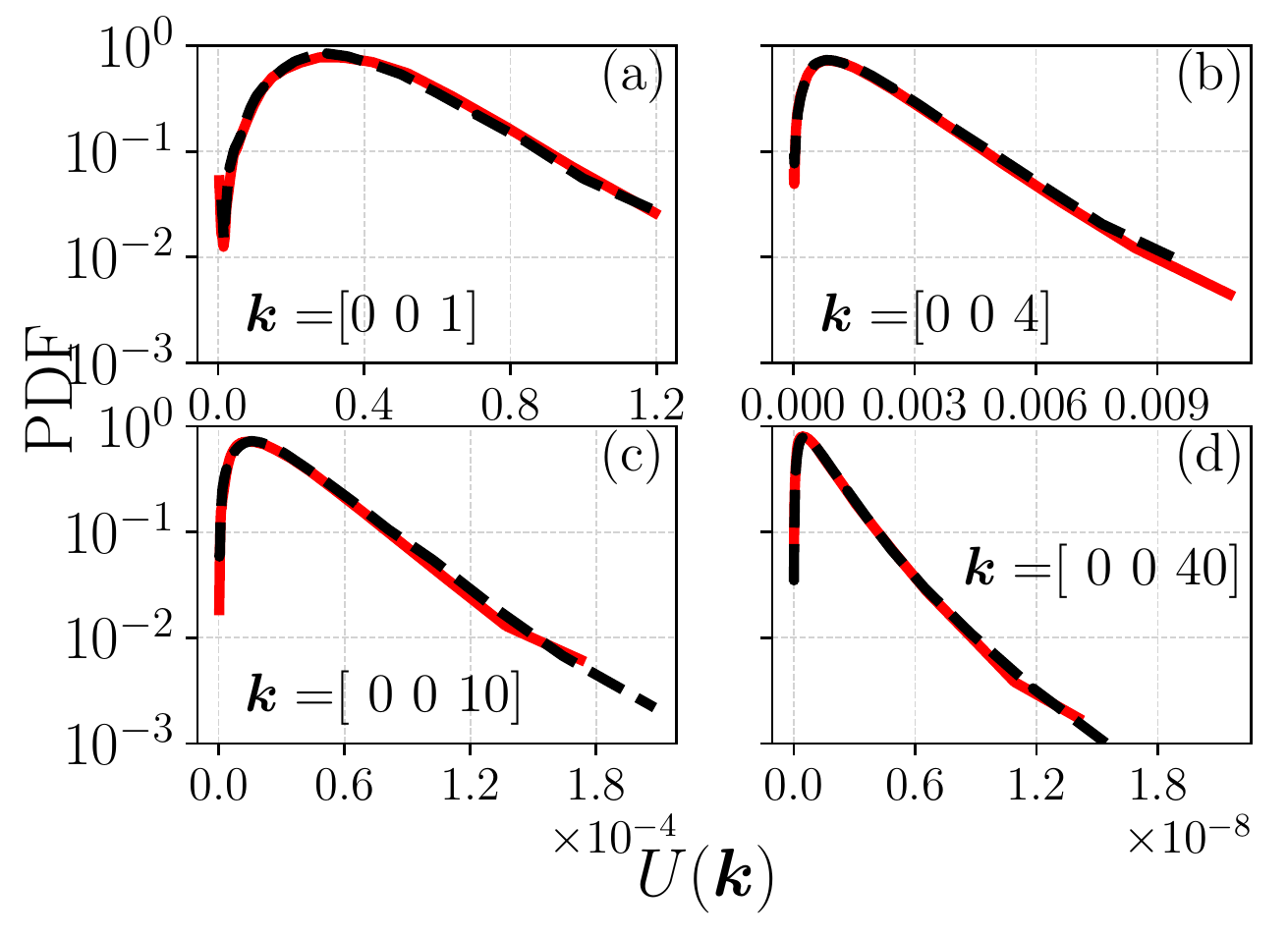}
		\caption{Probability distribution function of single mode kinetic energy $U(\kk)$ for chosen modes $\kk= (0,0,m)$, $m=1, \,4, \,10, \,40$, showing the equivalence between $\R$ (red straight lines) and $\I$ (black dashed lines). Here $\nu=10^{-2}$ and $N_0=128$ (R$7\bigcirc$), belonging to the hydrodynamic regime. }
		\label{fig:modes_equiv_conj2} 
	\end{figure}
	
	\begin{figure}[!h]
		\centering	
		\includegraphics[width=.85\linewidth]{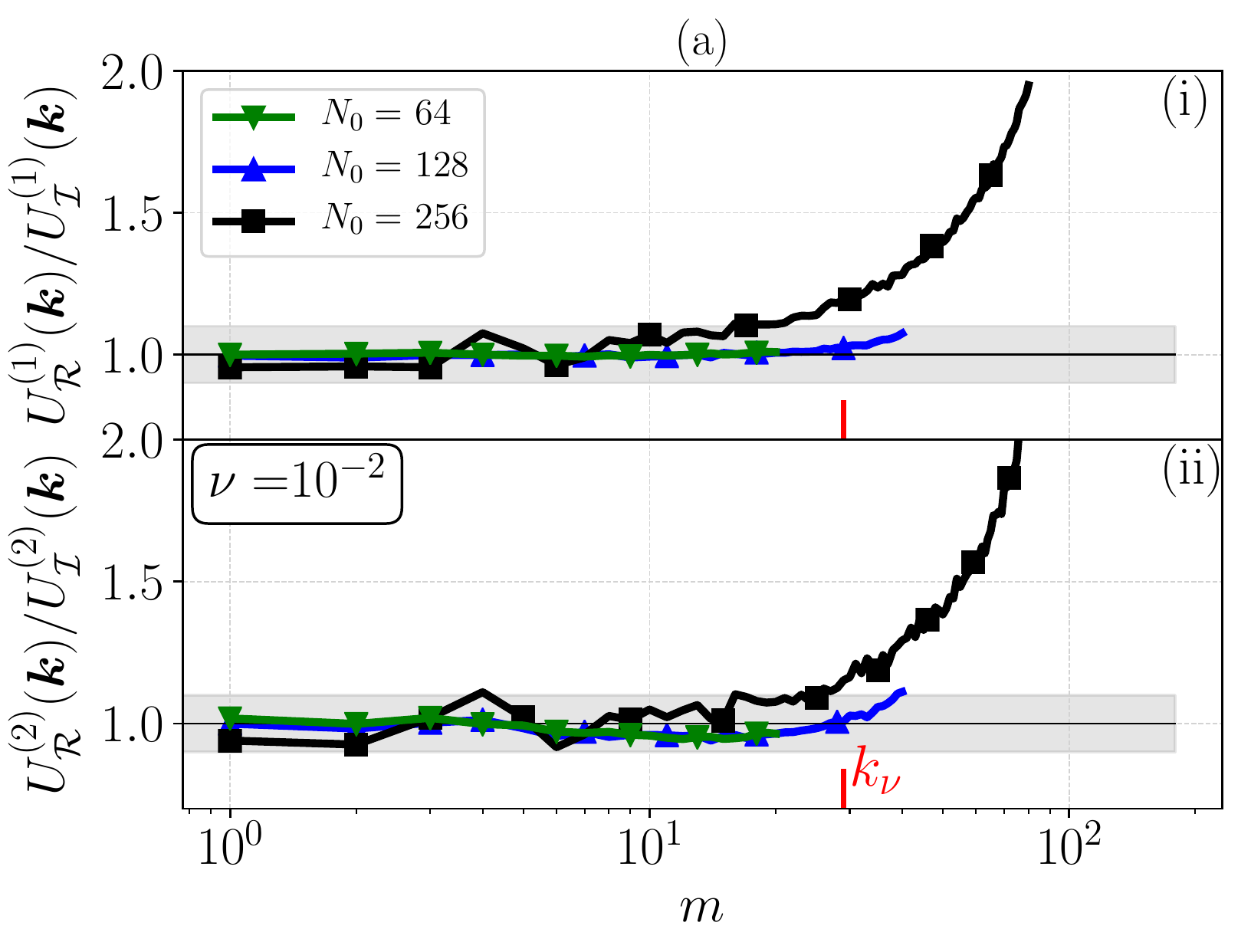}	
		\includegraphics[width=.85\linewidth]{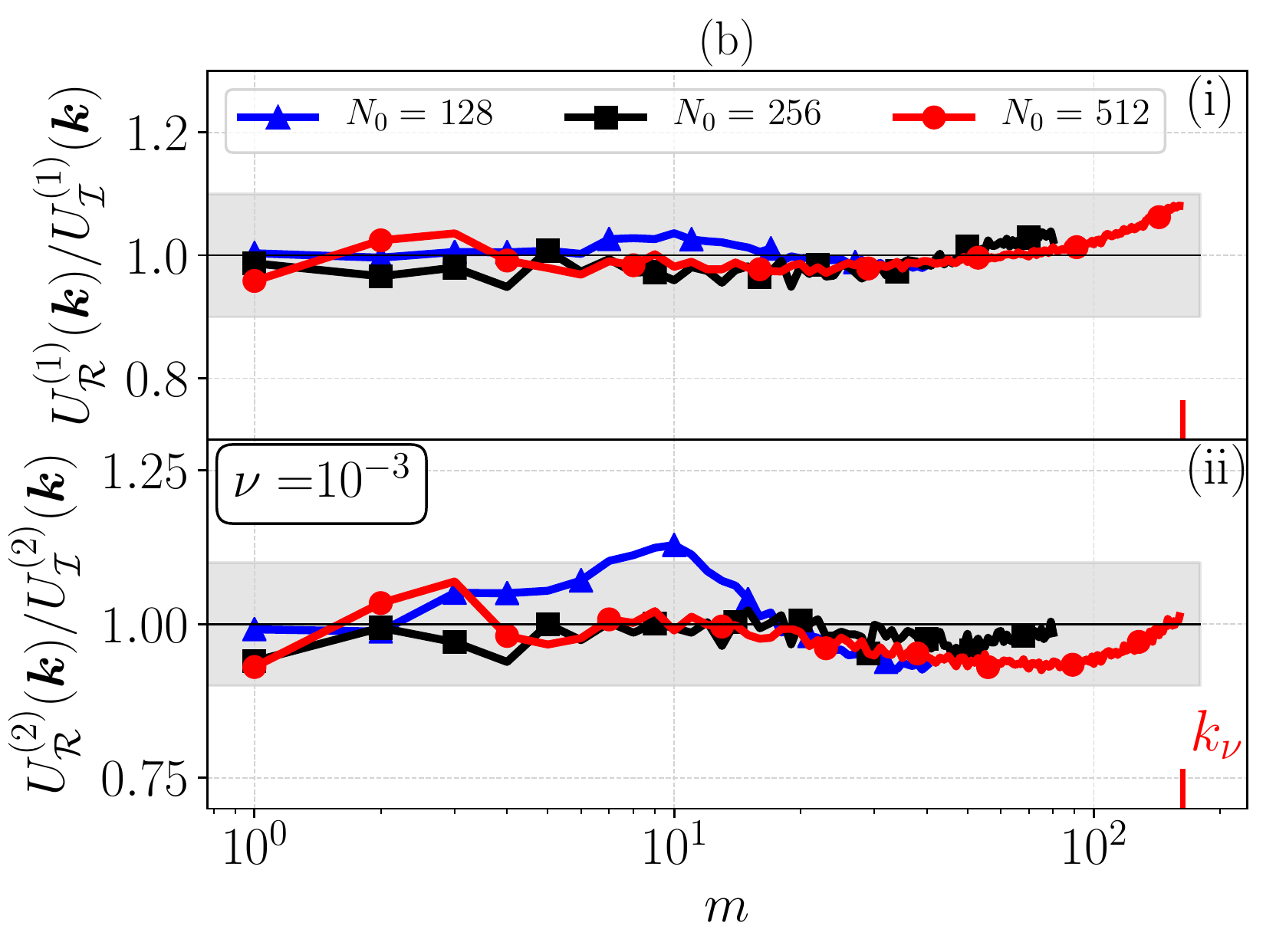}	
		\caption{Test of conjecture 2 for single modes kinetic energy $U(\kk)$. (i) Averaged $\R/\I$ ratio of mean kinetic energy of considered modes and (ii) standard deviation ratio, at (a) $\nu=10^{-2}$, and (b) $\nu=10^{-3}$. The gray band indicates a 10\% deviation from 1. The red vertical bar indicates the order of magnitude of the Kolmogorov scale, estimated as $k_\nu = \left(\frac{\varepsilon}{\nu^3}\right)^{1/4}$. }
		\label{fig:modes_equiv_ratios} 
	\end{figure}
	
	In Fig.~\ref{fig:modes_equiv_conj2} we fix $N_0=128$, $\nu=10^{-2}$ (R$7\bigcirc$) and compare the PDF of single mode kinetic energy, shifted to their mean, for the modes $\kk=(0,0,m)$, $m=1,$ 4, 10, 40 between $\R$  and $\I$. The agreement between the two ensembles is compelling. 
	
	A further analysis can be seen in Fig.~\ref{fig:modes_equiv_ratios} where all mean and standard deviation ratios of the time series have been collected for (a) $\nu=10^{-2}$  and (b) $\nu=10^{-3}$ for different values of $N$ and $m$. In Fig.~\ref{fig:modes_equiv_ratios}, for scales beyond the Kolmogorov scale $k_\nu$ (indicated by the red tick), we observe a disagreement between $\R$ and $\I$.

	Therefore, one could argue that, within the numerical set-up we have investigated in this work,  $k_\nu$ is an approximate upper bound (in the sense of order of magnitude) for the maximum wave number $K$ for which the conjecture 2 applies; or as stated after Eq.~\eqref{e2.6}, $c_\nu \approx 1$, both when considering $U^{(1)}(\kk)$, and $U^{(2)}(\kk)$. The observed difference between the $\R$ and $\I$ ensembles for $k \gtrsim k_\nu$ is further confirmed for all our well-resolved runs in the hydrodynamic regime. On the other hand, in the cross-over regime, \ie, here for R$2\bigcirc$ in Fig.~\ref{fig:modes_equiv_ratios}(a) and R$8\triangle$, R$13\triangle$ in Fig.~\ref{fig:modes_equiv_ratios}(b), the agreement for single mode statistics of the two ensembles, although not expected by conjecture 2, is satisfactory.

	\subsubsection{Energy spectrum}

\begin{figure}[!h]
		\centering
		\includegraphics[width=1.0\linewidth]{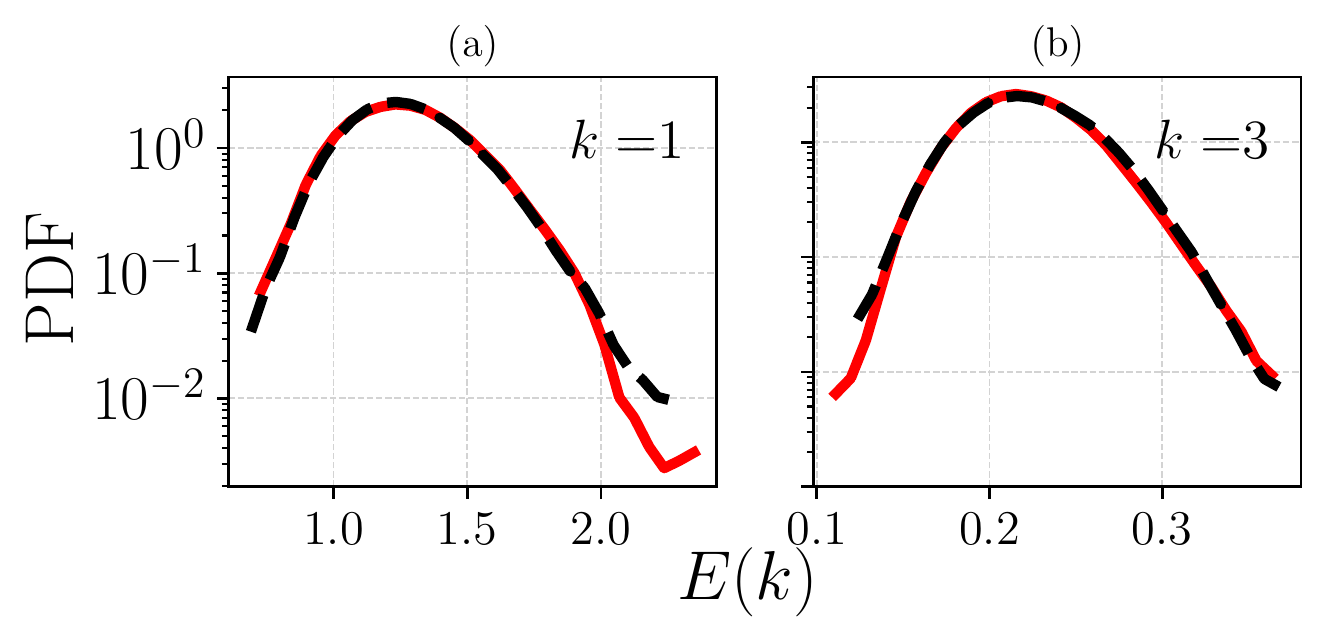}
		\caption{Probability distribution function of energy spectrum showing equivalence between $\R$ (red straight lines) and $\I$ (dashed black lines), for $k=$ 1, 3. For (a) and (b) it holds $E^{(1)}_{\R} (k)/E^{(1)}_{\I} (k)\approx 1$, and $E^{(2)}_{\R} (k)/E^{(2)}_{\I} (k)\approx 1$, confirming full equivalence between $\R$ and $\I$ up to $k \sim 4$ with respect to $E(k)$ at $\nu=10^{-2}$ Here $\nu=10^{-2}$ and $N_0=128$ (R7$\bigcirc$) corresponding to the hydrodynamic regime, and the $y$ axis is in logarithmic scale. }
		\label{fig:hist_shells_equiv_hydr_nu1e-2} 
	\end{figure}

\begin{figure}[!h]
		\centering
		\includegraphics[width=1.0\linewidth]{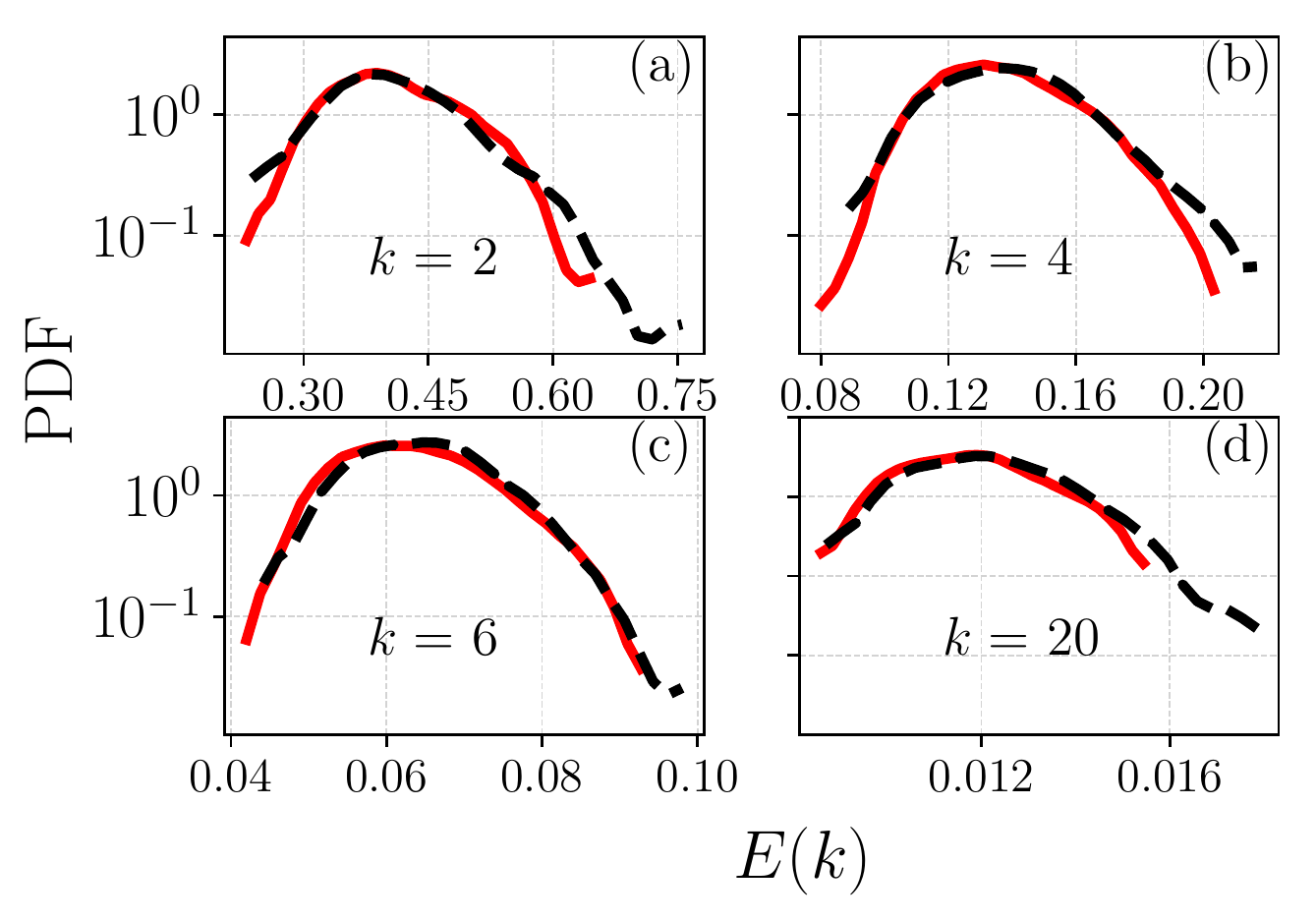}
		\caption{Probability distribution function of energy spectrum showing equivalence between $\R$ (red straight lines) and $\I$ (dashed black lines), for $k=$ 2, 4, 6, 20. For (a-d) it holds $E^{(1)}_{\R} (k)/E^{(1)}_{\I} (k)\approx 1$, and $E^{(2)}_{\R} (k)/E^{(2)}_{\I} (k)\approx 1$, confirming full equivalence between $\R$ and $\I$ up to $k\sim20$ with respect to $E(k)$ at $\nu=10^{-3}$. Here $N_0=512$ and $\nu=10^{-3}$ (R16$\bigcirc$) corresponding to the hydrodynamic regime, and the $y$ axis is in logarithmic scale. }
		\label{fig:hist_shells_equiv_hydr_nu1e-3} 
	\end{figure}

	With respect to the energy spectrum at the hydrodynamic regime, we first present the distributions at low $k$. In Fig.~\ref{fig:hist_shells_equiv_hydr_nu1e-2} we plot the PDF of energy spectra at $k=1$ and 3 for $\nu=10^{-2}$ at $N_0=128$ (R7$\bigcirc$) and we observe a very good agreement between $\R$ (red line) and $\I$ (black dashed line). Then, in Fig.~\ref{fig:hist_shells_equiv_hydr_nu1e-3} the same are shown for a larger range of $k$, \ie, $k=$2, 4, 6, 20 and a smaller viscosity, $\nu=10^{-3}$ at $N_0=512$ (R16$\bigcirc$), which is the only fully resolved run at $\nu=10^{-3}$. Again, the agreement between the two ensembles is remarkable, and slight disagreement at the tails is likely due to statistical accuracy.
	
	\begin{figure}[h]
		\centering	
		\includegraphics[width=1.\linewidth]{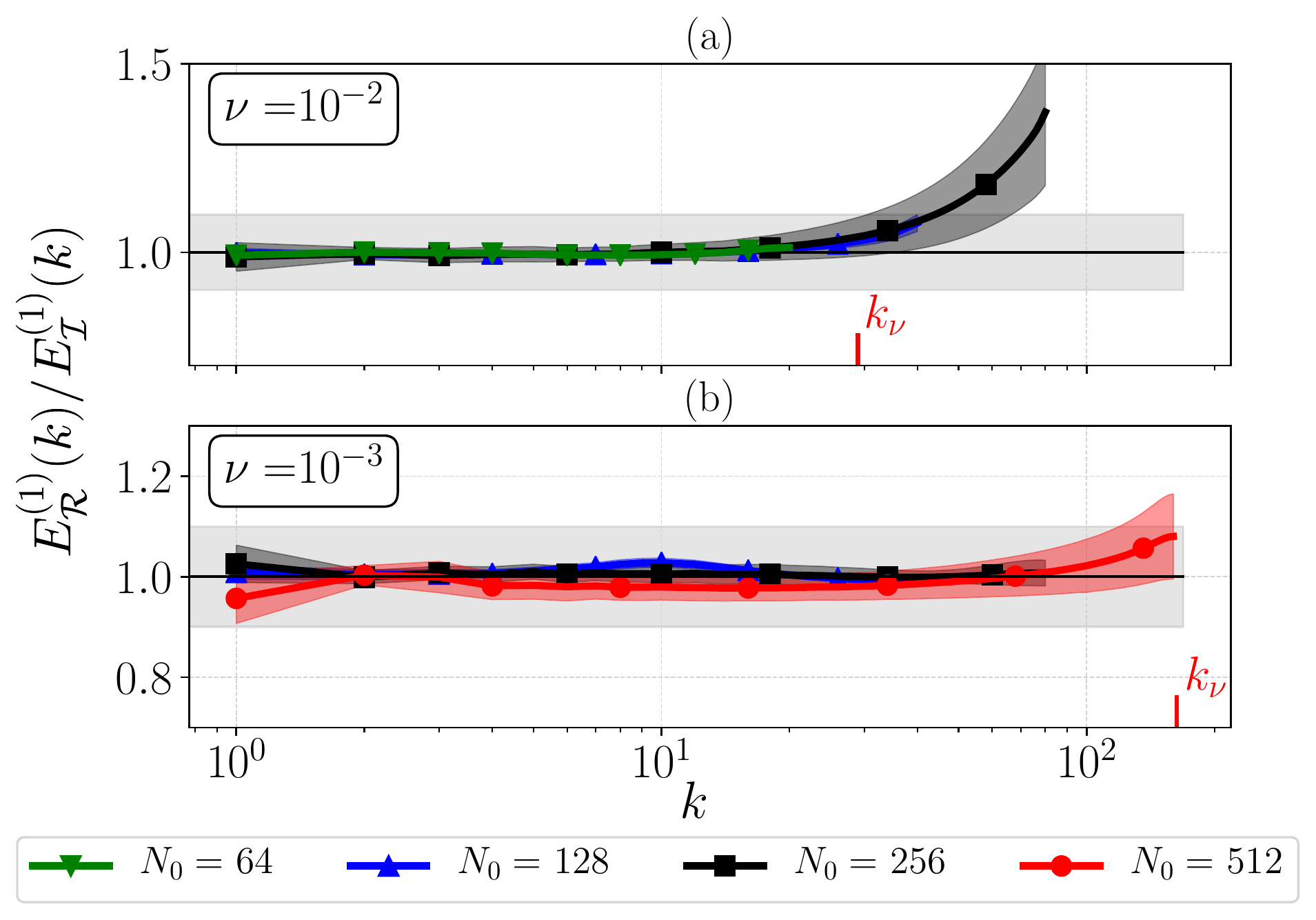}
		\caption{Test of conjecture 2 for the energy spectrum, considering the $\R/\I$ ratio of its mean at shell $k$ for $k=1,\ldots,N$, at (a) $\nu=10^{-2}$, and (b) $\nu=10^{-3}$. The transparent region corresponds to the statistical error. The gray band indicates a 10\% deviation from 1. $k_\nu$ shows the location of the Kolmogorov scale. The plots share the same $x$ axis. }
		\label{fig:shells_equiv_ratios_hdr_mean} 
	\end{figure}

	In Fig.~\ref{fig:shells_equiv_ratios_hdr_mean} we collect the mean energy spectrum $E^{(1)}(k)$ at different $k$. We test the conjecture 2 by quantifying the agreement between $\R$ and $\I$ through the ratio $E_{\R}^{(1)}(k) / E_{\I}^{(1)}(k)$, while fixing $\nu$, at changing $N$. A deviation from 1 would indicate disagreement with respect to this observable. In Fig.~\ref{fig:shells_equiv_ratios_hdr_mean}(a) at $\nu=10^{-2}$ we remark that $N_0=64$ (R3$\triangle$) is slightly under-resolved (not well developed fall-off region in the spectrum), but still follows the same trend as the rest of the well resolved simulations at increasing $k$. Deviations occur for $k\gtrsim k_\nu$ when the ratio of $E^{(1)}(k)$ is considered. In Fig.~\ref{fig:shells_equiv_ratios_hdr_mean}(b) we show the results for $\nu=10^{-3}$, where for all cases the ratio is close to 1 up until $k\sim k_\nu$, which is close to the cut-off of $N_0=512$. Also in (b),  the ratio $E_{\R}^{(1)}(k) / E_{\I}^{(1)}(k)$ of the under-resolved runs ($N_0=128$ and $256$) stays unity for all $k$.

	\subsubsection{Empirical determination of the locality cutoff $K$}\label{sec:locality_k}
	
	So far, the cases we have studied reveal a very good agreement between $\R$ and $\I$ for the quasithermalized regime (see Sec.~\ref{sec:Conj1_test}), which confirms conjecture 1, and full agreement for the hydrodynamic regime for $k\lesssim k_\nu$ when considering $U^{(1)}(\kk)$, $U^{(2)}(\kk)$, and $E^{(1)}(k)$. In accordance with the definition of conjecture 2, we recall that a scale $K$ can be defined such that observables depending on $\uu_\kk$ with $k< K$ can be considered ``local'', thus determining the domain of validity of conjecture 2. Furthermore, conjecture 2 states that $K<c_\n k_\n$ with $0<c_\n\to c_0\le\infty$ as $\n \to 0$. We state now that $c_\nu$ and accordingly $K$, can differ when considering different observables for testing the equivalence. Indeed, for $U^{(1)}(\kk)$, $U^{(2)}(\kk)$, and $E^{(1)}(k)$ we concluded that $c_\nu \sim 1$.
	
	\begin{figure}[b]
		\centering
		\includegraphics[width=1.0\linewidth]{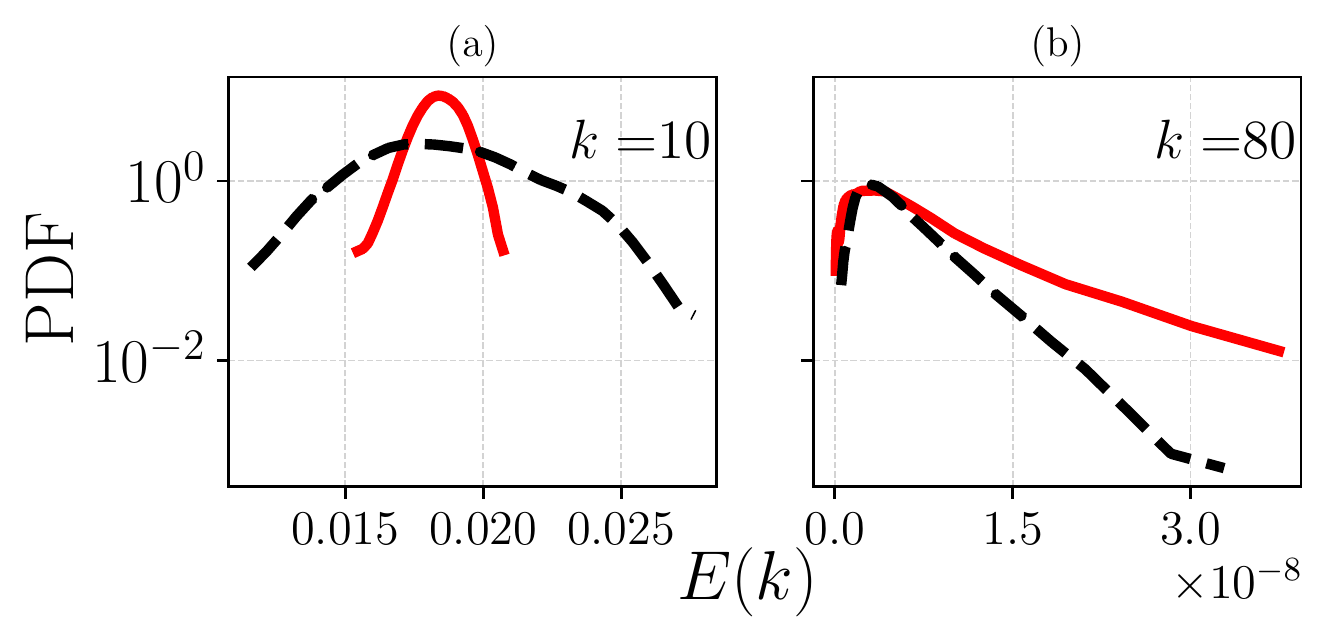}
		\caption{Probability distribution function of energy spectrum for $\R$ (red straight lines) and $\I$ (dashed black lines) ($k=$ 10 and $80>k_{\nu}\sim 29$). For $k=10$ the two distributions have the same mean. Here $\nu=10^{-2}$ and $N_0=256$ (R12$\bigcirc$) corresponding to the hydrodynamic regime, and the $y$ axis is in logarithmic scale. }
		\label{fig:hist_shells_equiv_hydr_nu1e-2_dis} 
	\end{figure}
	
	Next, we look at the statistics of energy spectrum at larger $k$. In Fig.~\ref{fig:hist_shells_equiv_hydr_nu1e-2_dis} we show two examples where disagreement is observed in the case of $\nu=10^{-2}$. Taking $N_0=256$ (R12$\bigcirc$) we show in Fig.~\ref{fig:hist_shells_equiv_hydr_nu1e-2_dis}(a) the PDF of $E(k)$ at $k=10$, for which the ratios $E^{(1)}_{\R} (k)/E^{(1)}_{\I} (k)\approx 1$ [see Fig.~\ref{fig:shells_equiv_ratios_hdr_mean}(a)], but $E^{(2)}_{\R} (k)/E^{(2)}_{\I} (k)\neq1$. In Fig.~\ref{fig:hist_shells_equiv_hydr_nu1e-2_dis}(b) we choose a $k$ larger than the Kolmogorov scale, \ie, $k=80 > k_{\nu} \sim 29$, and both ratios of mean and standard deviation between $\R$ and $\I$ are not equal to 1.

\begin{figure}[t]
		\centering	
		\includegraphics[width=1.\linewidth]{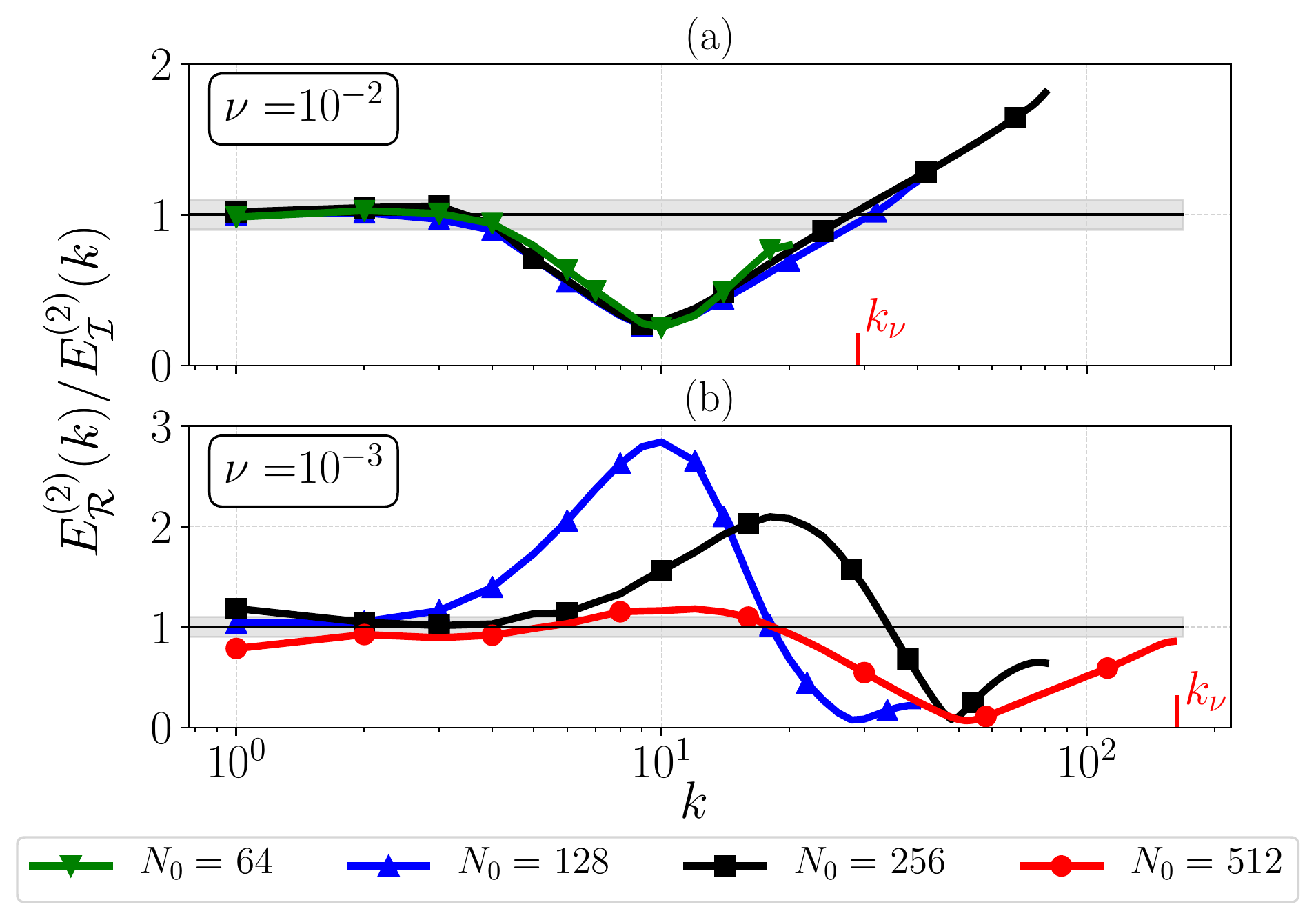}
		\caption{ Test of conjecture 2 for the energy spectrum, considering the $\R/\I$ ratio of its standard deviation at shell $k$ for $k=1,\ldots,N$, at (a) $\nu=10^{-2}$, and (b) $\nu=10^{-3}$. In (a) the runs with $N_0=128$ and $N_0=256$ are fully resolved and belong to the hydrodynamic regime, while in (b) only $N_0=512$ is fully resolved. The gray band indicates a 10\% deviation from 1. $k_\nu$ shows the location of the Kolmogorov scale. The range of validity of conjecture 2 increases as we decrease viscosity in the hydrodynamic regime, as expected. The plots share the same $x$ axis. }
		\label{fig:shells_equiv_ratios_hdr_std} 
	\end{figure}

	In Fig.~\ref{fig:shells_equiv_ratios_hdr_std} we quantify the behavior of the standard deviation $E^{(2)}(k)$ of each ensemble, by considering their ratio $\R/\I$ for all $k$. conjecture 2 is tested for fixed $\nu$, (a) $\nu=10^{-2}$ and (b) $\nu=10^{-3}$ at increasing $N$. Starting from $k=1$ there is initially agreement between $\R$ and $\I$ but we observe that $E^{(2)}_{\R} (k)/E^{(2)}_{\I} (k)$ departs from unity at a smaller $k$ than $E^{(1)}_{\R} (k)/E^{(1)}_{\I} (k)$, which is also smaller than $k_\nu$. Notice that in (a) the runs with $N_0=128$ and $256$ are fully resolved and belong to the hydrodynamic regime, and although the runs at $N_0=64$ are slightly under-resolved, it follows the same trend with the rest. Moreover, once the cut-off $N$ is larger than $k_\nu$ by a sufficient margin, further increases in $N$ do not change the features of the ratio curves, both for $E^{(1)}_{\R} (k)/E^{(1)}_{\I} (k)$ and $E^{(2)}_{\R} (k)/E^{(2)}_{\I} (k)$. Instead, in (b) only the $N_0=512$ case is fully resolved, and although $E^{(1)}_{\R} (k)/E^{(1)}_{\I} (k)\approx 1$ for all $N_0$ [see Fig.~\ref{fig:shells_equiv_ratios_hdr_mean}(b)] the standard deviation ratios differ when changing $N_0$. This occurs because $N_0=128$ and $256$ are under-resolved at $\nu=10^{-3}$, and belong to the crossover regime (hence outside the validity of conjecture 2), but are still displayed to show the transition from crossover to hydrodynamic regime, and how the agreement improves in that case.

	To be more specific about the deviation from 1 with respect to Fig.~\ref{fig:shells_equiv_ratios_hdr_std}, for (a) at $\nu=10^{-2}$ it appears that $K\approx 4$. Instead, from Fig.~\ref{fig:shells_equiv_ratios_hdr_std}(b) at $\nu=10^{-3}$ it appears $K\approx 20$. The fact that $K$ increases at decreasing $\nu$, with sufficiently large $N$, is foreseen by conjecture 2, and confirmed here. In other words, equivalence between $\R$ and $\I$ holds for a larger range of $k$, as $\nu$ decreases, when the testing criterion is the shape of the distributions of $E(k)$, or similarly the standard deviation of it, $E^{(2)}$. 
	
	Note that the location of the minimum of $E^{(2)}_{\R} (k)/E^{(2)}_{\I} (k)$ in Fig.~\ref{fig:shells_equiv_ratios_hdr_std} shifts with changing $\nu$ for well-resolved simulations. It actually corresponds to the end of the inertial range, and it is related to the chosen observable that is kept fixed in $\R$. Since here we chose the enstrophy $\DD(\uu)$, which is dominated by the small scales, i.e., large $k$, the constraint $\media{\DD(\uu)}_\n^{\I,N}=D$ actually suppresses the fluctuations of $E_\R(k)$, for those $k$ around the end of the inertial range; see also Fig.~\ref{fig:hist_shells_equiv_hydr_nu1e-2_dis}(a). A similar result was observed in \cite{BCDGL018}. 
	
\begin{figure}[t]
		\centering	
		\includegraphics[width=0.9\linewidth]{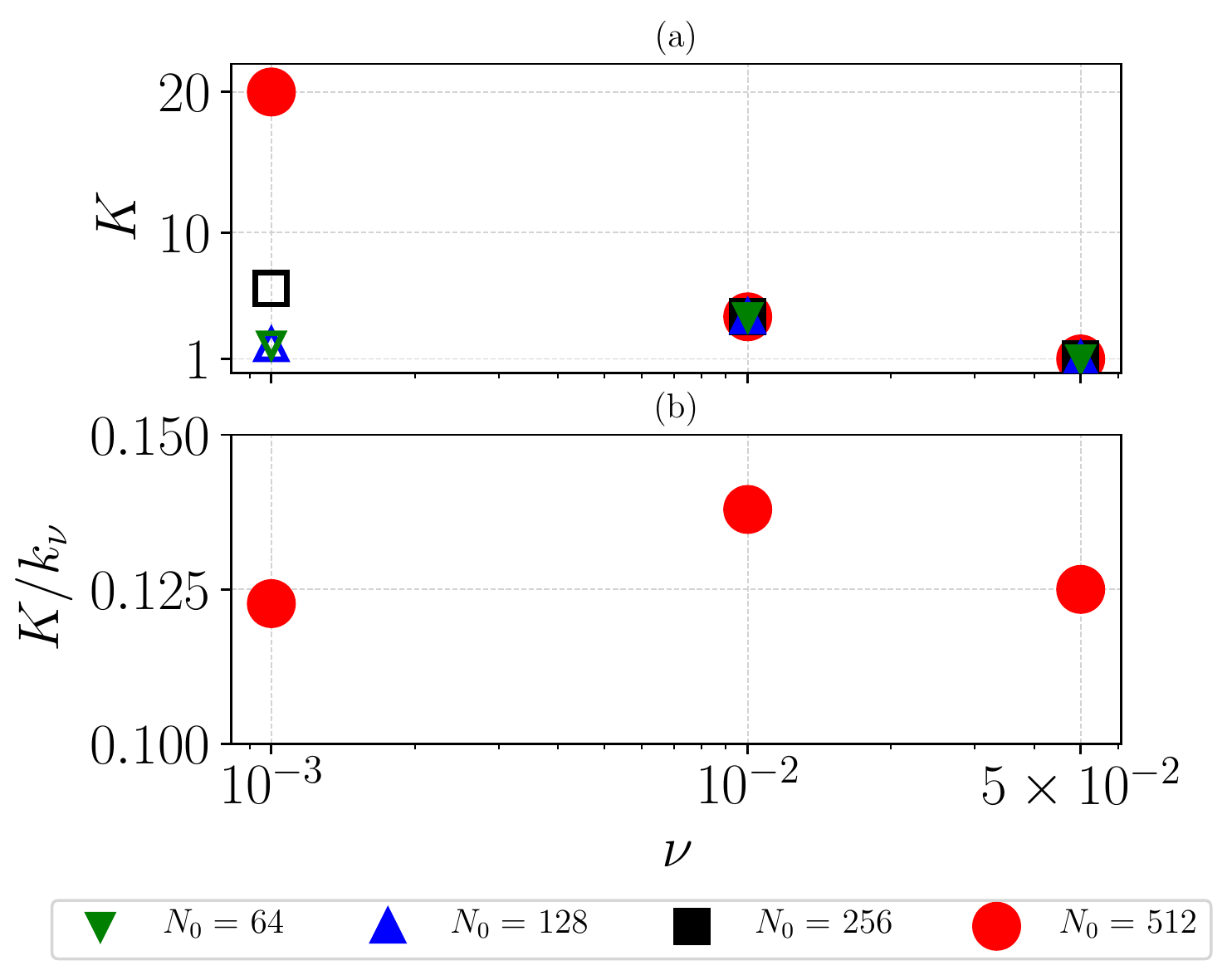}	
		\caption{(a) A quantitative estimate of the locality cut-off scale $K$, defined as the maximum wave number where all observables used to check the equivalence (conjecture 2)  have the same value within $10\%$, versus $\nu$ at different $N_0$.  Here filled points correspond to runs in the hydrodynamic regime and open points to the crossover regime. (b) The ratio $c_{\nu} = K/k_\nu$  for all runs in the hydrodynamic regime as a function of $\nu$, which is approximately constant. Both plots share the same $x$ axis in log scale. 	}
		\label{fig:KnuN_real} 
	\end{figure}

	The resulting values of $K$ as extracted from second order moments of the energy spectrum, $E^{(2)} (k)$, are gathered in Fig.~\ref{fig:KnuN_real}(a) for runs in the mixed hydrodynamic-crossover regime. Both at $\nu=5\times10^{-2}$ and $\nu=10^{-2}$ the runs are in the hydrodynamic regime, while at $\nu=10^{-3}$ only $N_0=512$ (R16$\bigcirc$) is well resolved, for which we notice that $K$ is maximum. In Fig.~\ref{fig:KnuN_real}(b) we show that the empirically determined $K$ does scale proportionally to $k_\nu$, indicating that the range of scales where conjecture 2 holds is increasing with the Reynolds number, i.e. supporting the statements that conjecture 2 is uniformly valid in a sub-set of the inertial range for all turbulent intensities.
	
	From Figs.~\ref{fig:modes_equiv_conj2}--\ref{fig:hist_shells_equiv_hydr_nu1e-2_dis}, as well as Fig.~\ref{fig:KnuN_real}(b), and with reference to conjecture 2 it appears that within  the case studied here we have the following:
	\begin{itemize}
		\item $c_\nu \approx 1$ when considering $U^{(1)}(\kk)$, $U^{(2)}(\kk)$, and $E^{(1)}(k)$ for the Equivalence test, and
		\item $c_\nu \approx 1/8$ when considering $E^{(2)}(k)$.
	\end{itemize} 
	All the above confirm the validity of conjecture 2 up to some wave number $K$.

	\begin{figure*}[t]
		\centering	
		\includegraphics[width=.43\linewidth]{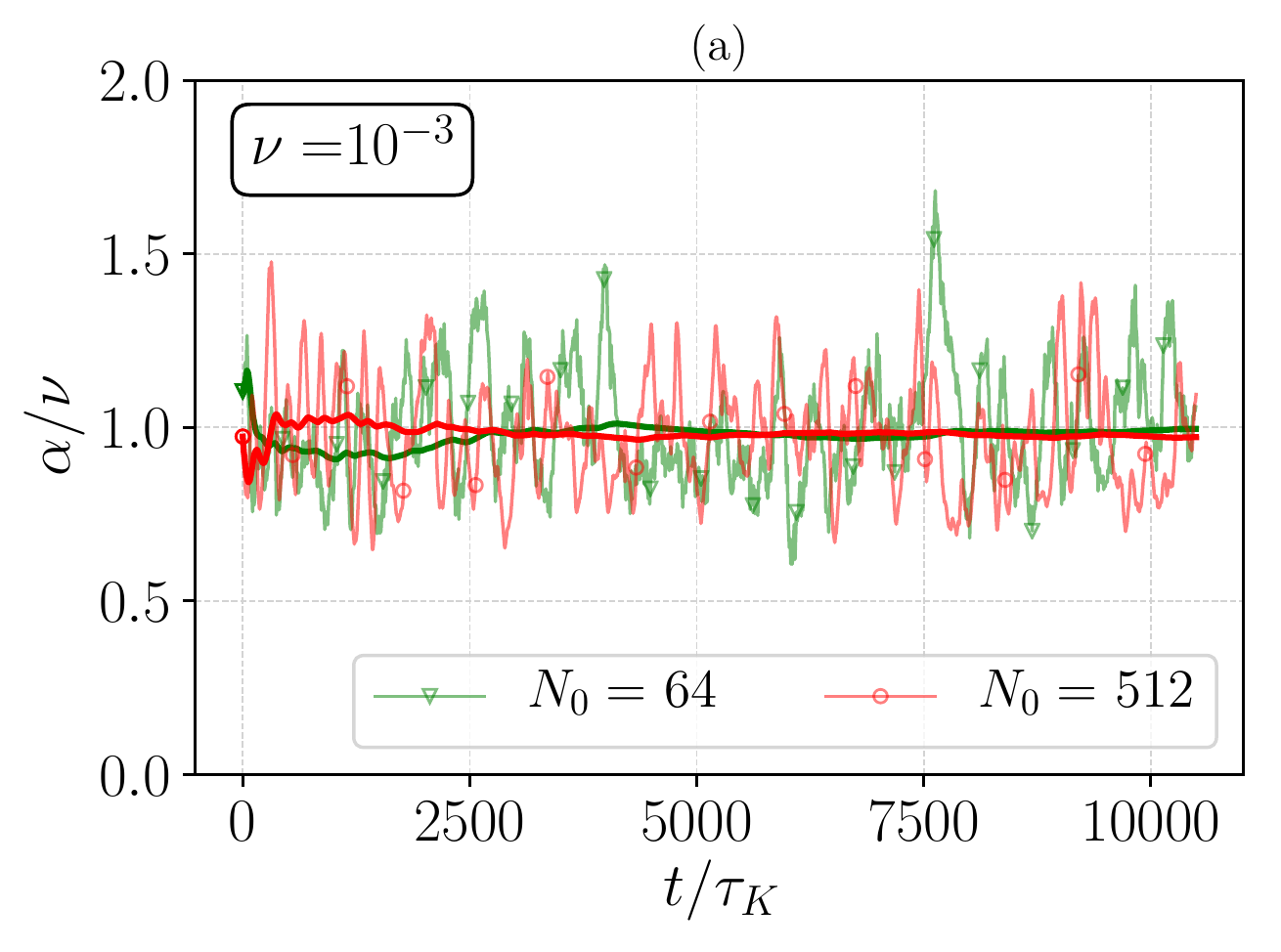}
		\includegraphics[width=.43\linewidth]{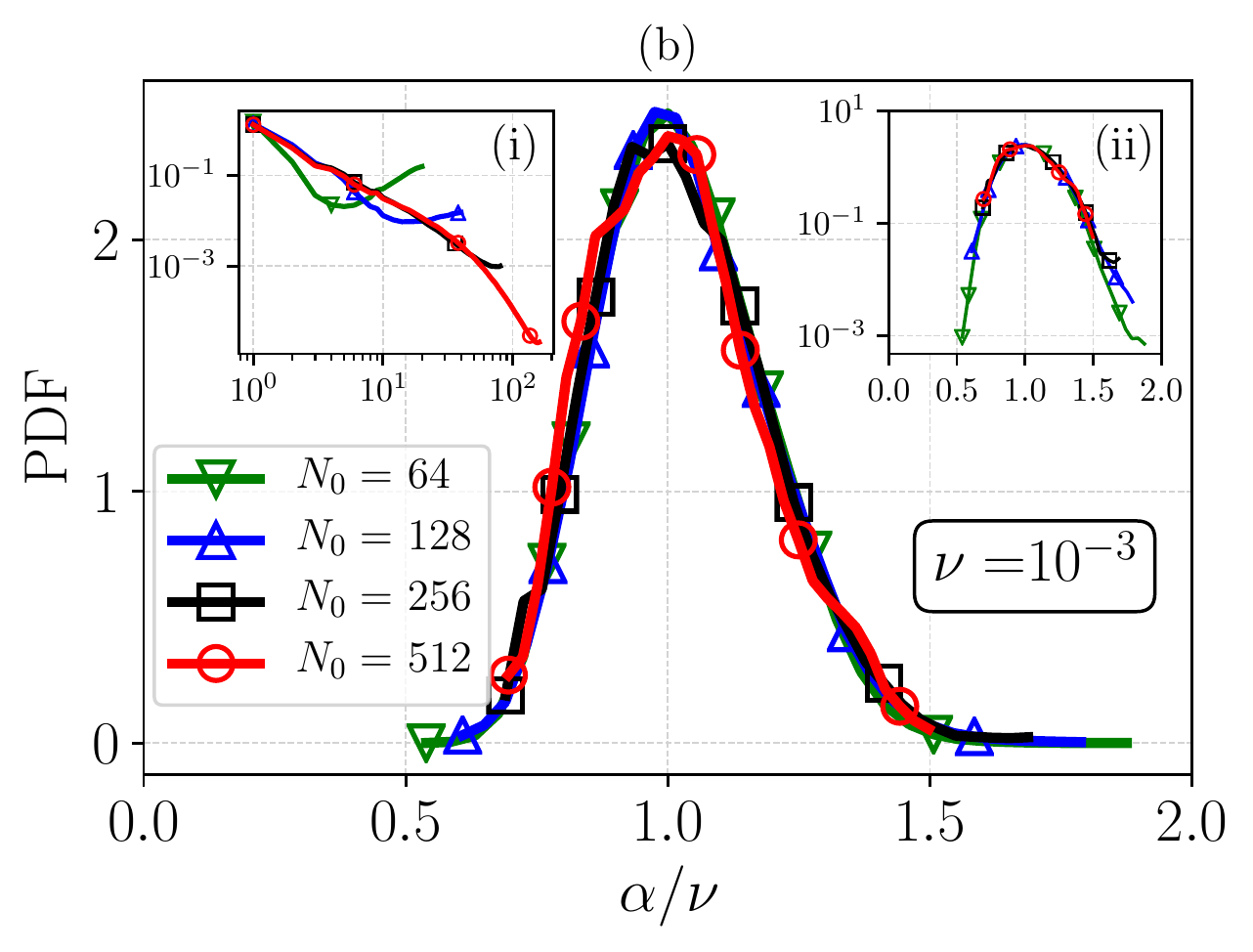}
		\includegraphics[width=.43\linewidth]{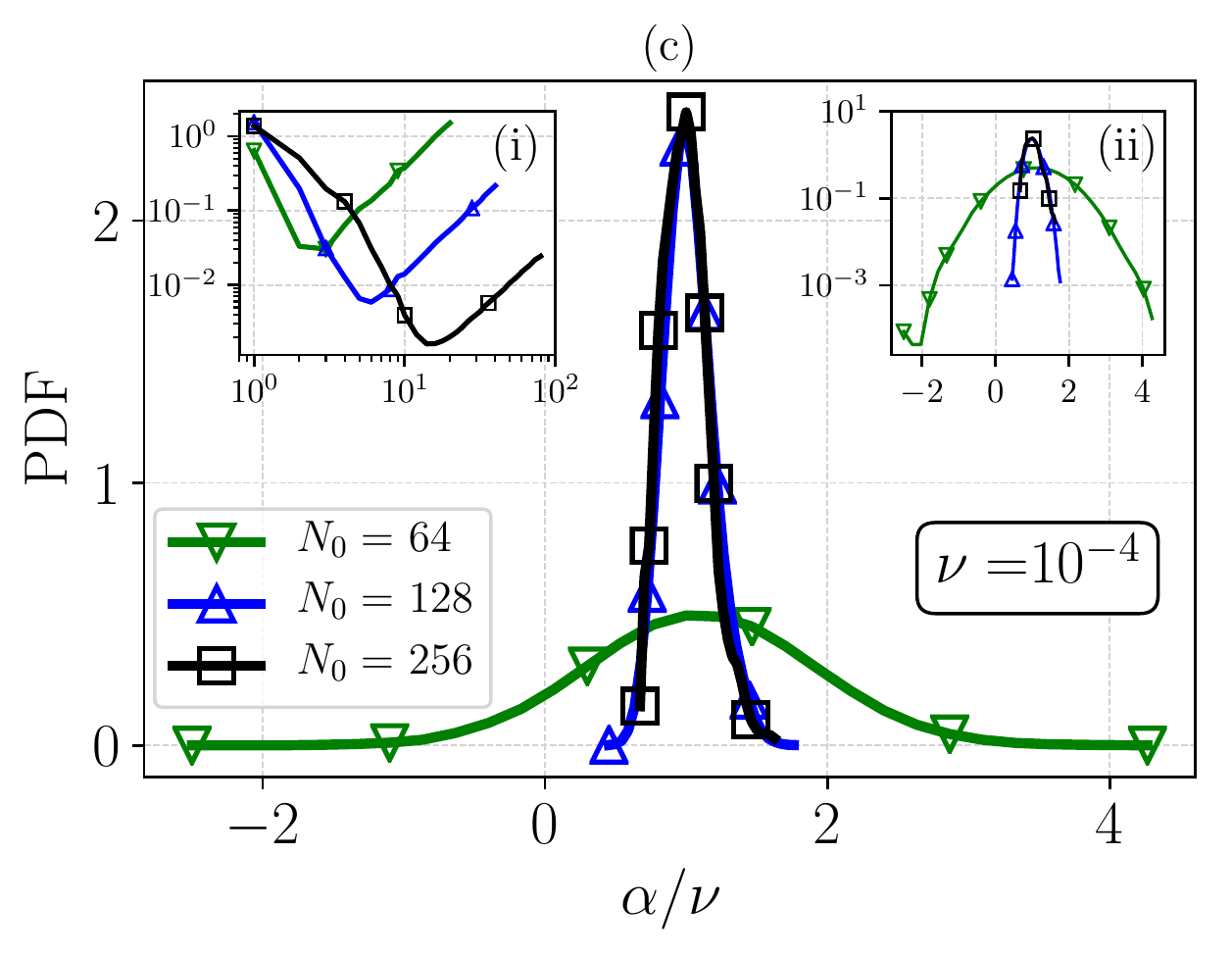}
		\includegraphics[width=.43\linewidth]{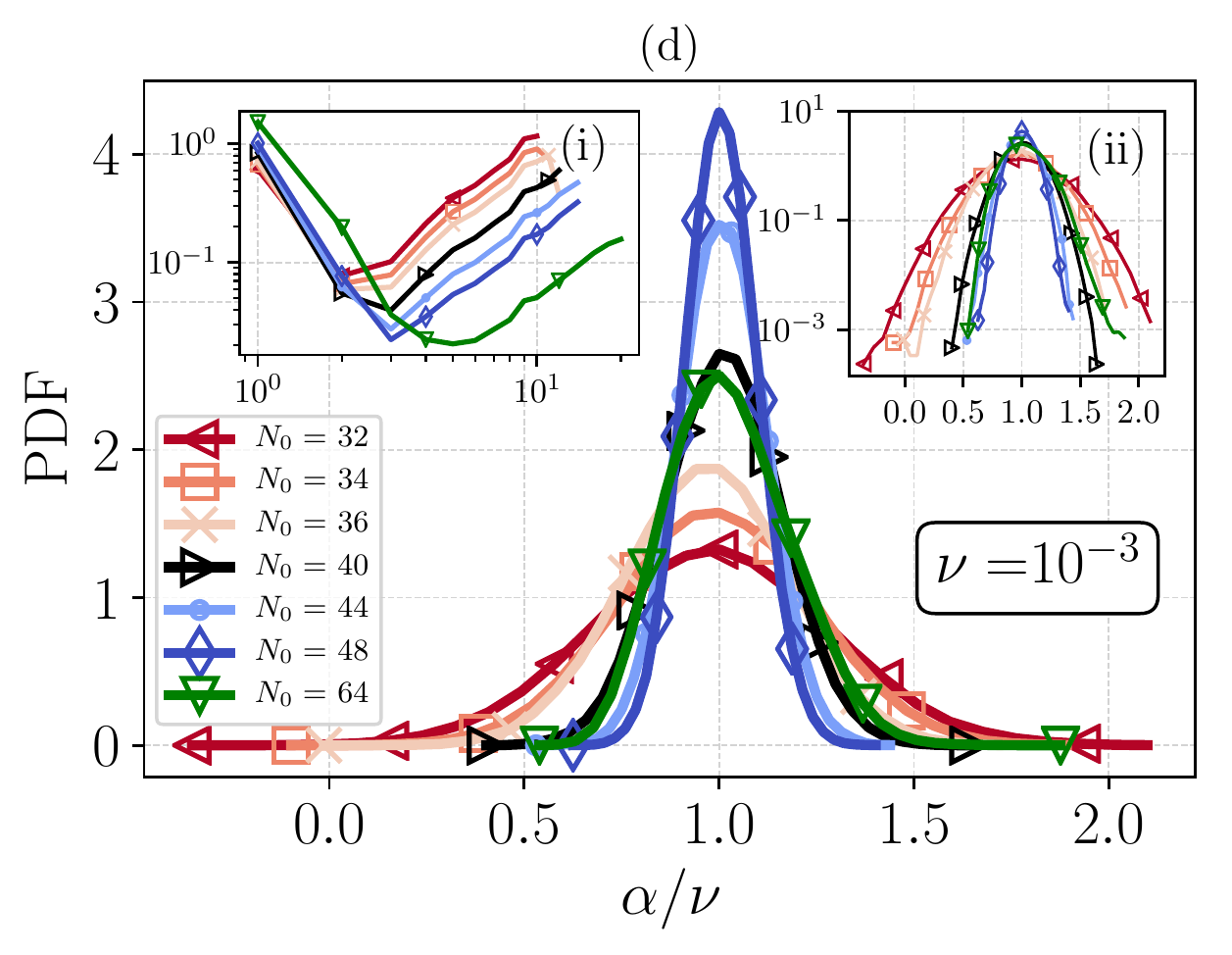}
		\caption{Statistics of $\alpha$ for $\R$. (a) Temporal evolution of $\alpha/\nu$ as a function of  time, which is rescaled by the Kolmogorov time scale $\tau_K$, at $\nu=10^{-3}$. In solid line, green (for $N_0=64$) and red (for $N_0=512$) is the running average of $\alpha/\nu$. The fact that $\langle \alpha/\nu \rangle = 1$ in $\R$ is a rigorous prediction of conjecture 2 [see also Eq.~\eqref{eq:fl_visc}]. (b) PDF of $\alpha/\nu$ obtained considering the whole available statistics for each $N_0$, at $\nu=10^{-3}$. The inset (i) is the energy spectrum, and inset (ii) is the PDF in logarithmic scale. (c) Same as (b) at $\nu=10^{-4}$. (d) Same as (b) at $\nu=10^{-3}$ showing successively small $N_0$ between 32 and 64. 	}
		\label{fig:alpha_fluct} 
	\end{figure*}

	\def\SEC{Properties of the Reversible Navier-Stokes equations}
	\section{\SEC}
	\label{sec4}
	In the previous section we have established the domain of validity of conjecture 2, and hence the equivalence of the two non-equilibrium ensembles, using a dimensionless scale-by-scale comparison  for different  local observable, based  on single Fourier modes energy or shell energy (see Figs.~\ref{fig:shells_equiv_ratios_conj1_qeq}, \ref{fig:modes_equiv_ratios}, \ref{fig:shells_equiv_ratios_hdr_mean}, and \ref{fig:shells_equiv_ratios_hdr_std}).  We have shown that the equivalence holds up to some wave number $K$. Similar results are shown when comparing  the whole probability distribution function of the same quantities. These results extend and clarify the recent investigations \cite{SDNKT018,JC020} where the equivalence was tested on the basis of the energy spectrum or on the global scaling properties of high order moments of the velocity increments in the inertial range only.
	
	Let us remark that the property valid in $\R$ that $\DD(\uu)$ is a finite non-zero constant of motion has obvious implications on the sequence of solutions $\uu^N(t)$ and for the Leray's solutions \cite{Le934,CKN982}, which are their weak limits as $N\to\infty$. Perhaps the most remarkable is an {\it a priori} upper bound on $\a(\uu)$ which can be proven (see Appendix \ref{appendixA1})  to satisfy, for any solution and any $N$, the condition:
	\be
	|\a(\uu)|<C_2 \left(\sqrt{D}+{\sqrt{D}}^{-1}\right).
	\Eq{eq:a_upper_bound}
	\ee
	
	A second interesting and potentially very important remark on the $\R$ evolution is that {\it if} for all $N$ large enough there were $\e>0$ such that also: $0<\e< \a<\k$, with probability $1$ in the stationary state, then, it is possible to conclude that the attractor consists of $\uu^N(t)$ with derivatives of all orders bounded uniformly in $N$, \ie, on the attractor the fields are uniformly smooth. This is an immediate consequence of the autoregularization results, (see Proposition 5, Sec. 3.2, in \cite{Ga002}, and \cite{Se962}), and of the constancy of the enstrophy $D$, as shown in the Appendix \ref{seca2}. Positivity of the lower bound on $\a$ is expected at large viscosity \ie, small $\re$.

	Note also that by combining the aforementioned bound of Eq.~\equ{eq:a_upper_bound}, as proposed in the literature, with the fact that the dissipation $\n D$ in $\I$ has a positive limit as $\n\to0$ should indicate, (see \cite[p.306]{Kr975a}), that the upper bound on $\a$ is $O(\n^{-\frac12})$, which puts a limit to the fluctuations of $\a$ (which by the conjecture is on average equal to $\n$; see also Sec.~\ref{sec:mean_alpha}). Furthermore the upper bound \equ{eq:a_upper_bound} suggests that the transport contribution to $\a$ (\ie, $\L(\uu)/\G$) might dominate  over the second term (\ie, the forcing contribution $(\int \D \ff\cdot\uu)/\G$), as confirmed in Fig.~\ref{fig:mean_alpha_R}(b).

	\subsection{Numerical results  about the sign of $\alpha$ in $\R$}
	\label{sec:num_res_alpha}
	
	In this section we inspect the statistics of the fluctuations of $\alpha(\uu)$ for $\R$, in order to assess the probability to observe negative values at varying the control parameters. In Fig.~\ref{fig:alpha_fluct}(a) we show in light colors the temporal evolution of $\a/\nu$, and in solid colors its running average, performed as $\frac{1}{t-t_0}\sum_{t'=t_0}^{t} \a(\uu(t'))/\nu$ for $\R$ at $\nu=10^{-3}$ and $N_0= 64, \,512$ (green -~R3$\triangle$ and red -~R16$\bigcirc$, respectively). The presented time window, which is scaled by the Kolmogorov time $\tau_K=\sqrt{\nu/\varepsilon}$, is the maximum achieved for the run with $N_0= 512$ (R16$\bigcirc$), while for R3$\triangle$ this is only about 1\% of its total temporal extend. Then, in Fig.~\ref{fig:alpha_fluct}(b) the PDFs of $\a/\nu$ are presented for $N_0=64$, 128, 256, and 512 at $\nu=10^{-3}$. Notably, all PDFs agree, even though at $\nu=10^{-3}$ only $N_0=512$ is fully resolved, belonging therefore to the hydrodynamic regime, while the other simulations belong to the crossover regime. The inset plots (i) and (ii) are respectively the energy spectrum and the PDF in semilogarithmic scale.

	In Fig.~\ref{fig:alpha_fluct}(c) we show the same of Fig.~\ref{fig:alpha_fluct}(b) but at lower viscosity, $\nu=10^{-4}$, where the data set at $N_0=64$ (R4$\square$) is in the quasithermalized regime, while the other runs are in the crossover regime.  As one can see, except for the case of the quasithermalized regime, in all other simulations we do not observe negative values of $\a$ within our statistical sample.
	
	Additionally, in Fig.~\ref{fig:alpha_fluct}(d) we perform a detailed study at fixed $\nu=10^{-3}$ and low resolution, in order to observe when, and how the transition to the occurrence of frequent negative values of $\alpha$ takes place. Starting from $N_0=64$ and gradually decreasing $N_0$, we observe that the PDF of $\alpha$ first turn from non-Gaussian to quasi-Gaussian, while becoming narrower (see $N_0=48$ and $N_0=44$), before starting to widen (see $N_0=40$), until occurrence of $\a<0$ events appears (see $N_0\leq36$). The rapid change in the left tail of the PDF at changing $N_0$  suggests that in order to further substantiate the asymptotic probability to observe negative $\a$ values, when $N_0$ is large and  in the presence of intermediate or hydrodynamical regimes, one would need statistical samples significantly (exponentially) larger than those we could generate here, which is by far out of the scope of this paper and probably out of reach given the available computational power. This result is in agreement with some previous observations made in the context of a reduced model of turbulence \cite{BCDGL018}.

	In \cite{JC020}  negative values of $\alpha$ are observed in an reversible ensemble at $N_0=1024$ and Taylor-based Reynolds number Re$_\lambda=300$. Indeed, such results are somewhat puzzling and are not expected on the basis of our data. We do not expect  to be able to observe $\a<0$ at such large $N_0$ and such Reynolds number by extrapolating from our results for the fully resolved hydrodynamical regime  (see Sec.~\ref{sec:num_res_alpha_I} below).

	\begin{figure*}[t]
		\centering	
		\includegraphics[width=.43\linewidth]{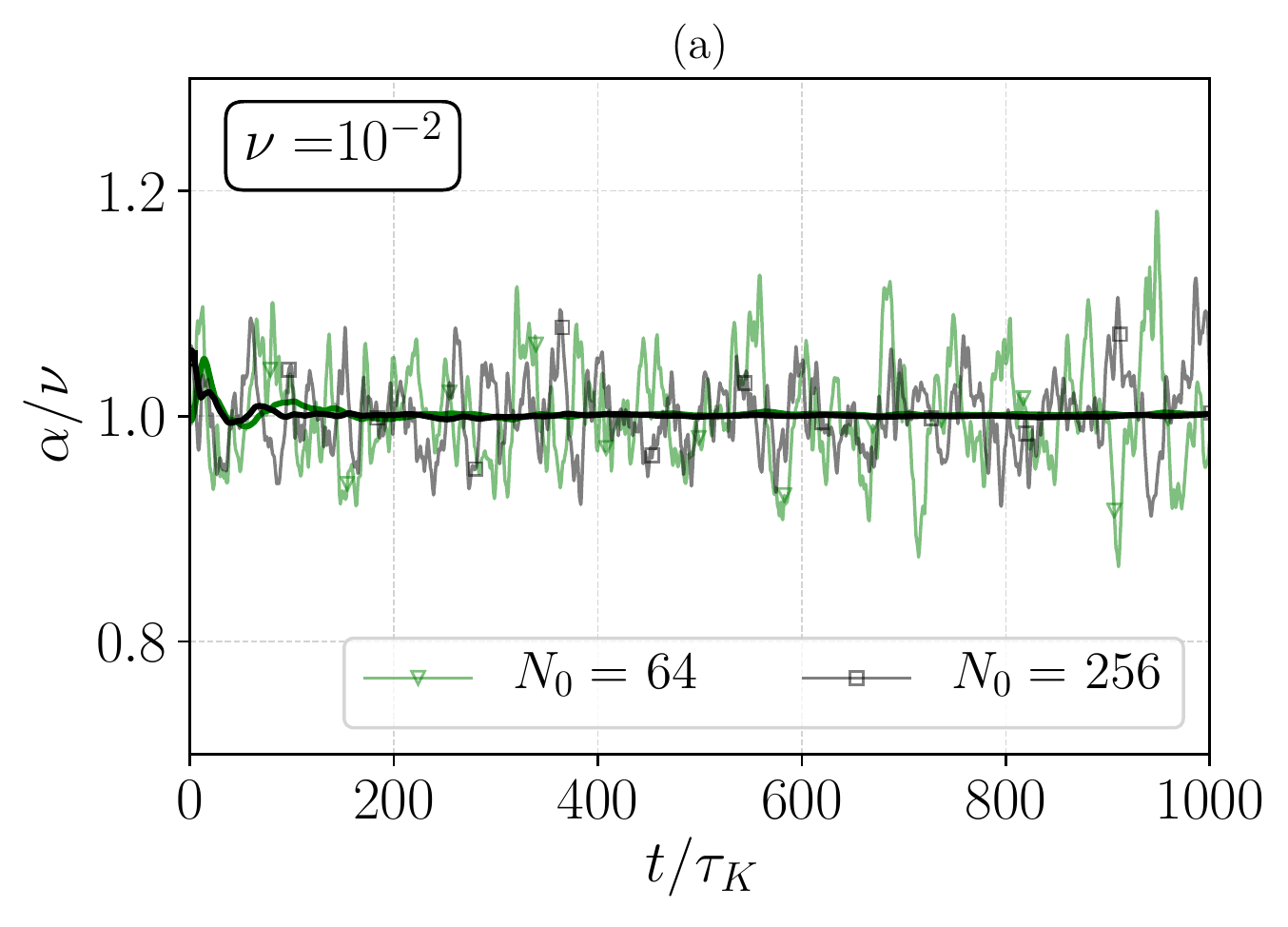}
		\includegraphics[width=.43\linewidth]{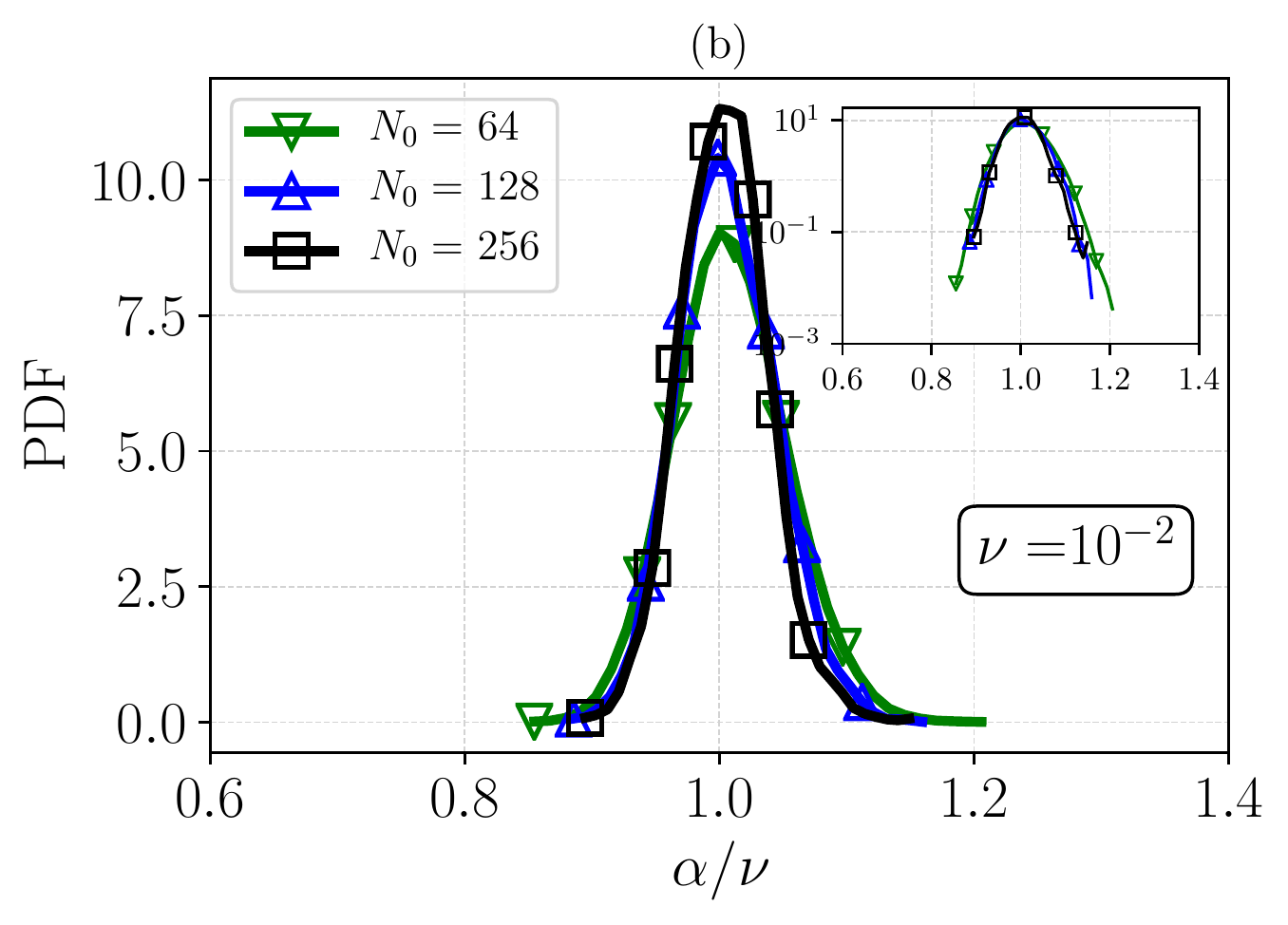}
		\includegraphics[width=.43\linewidth]{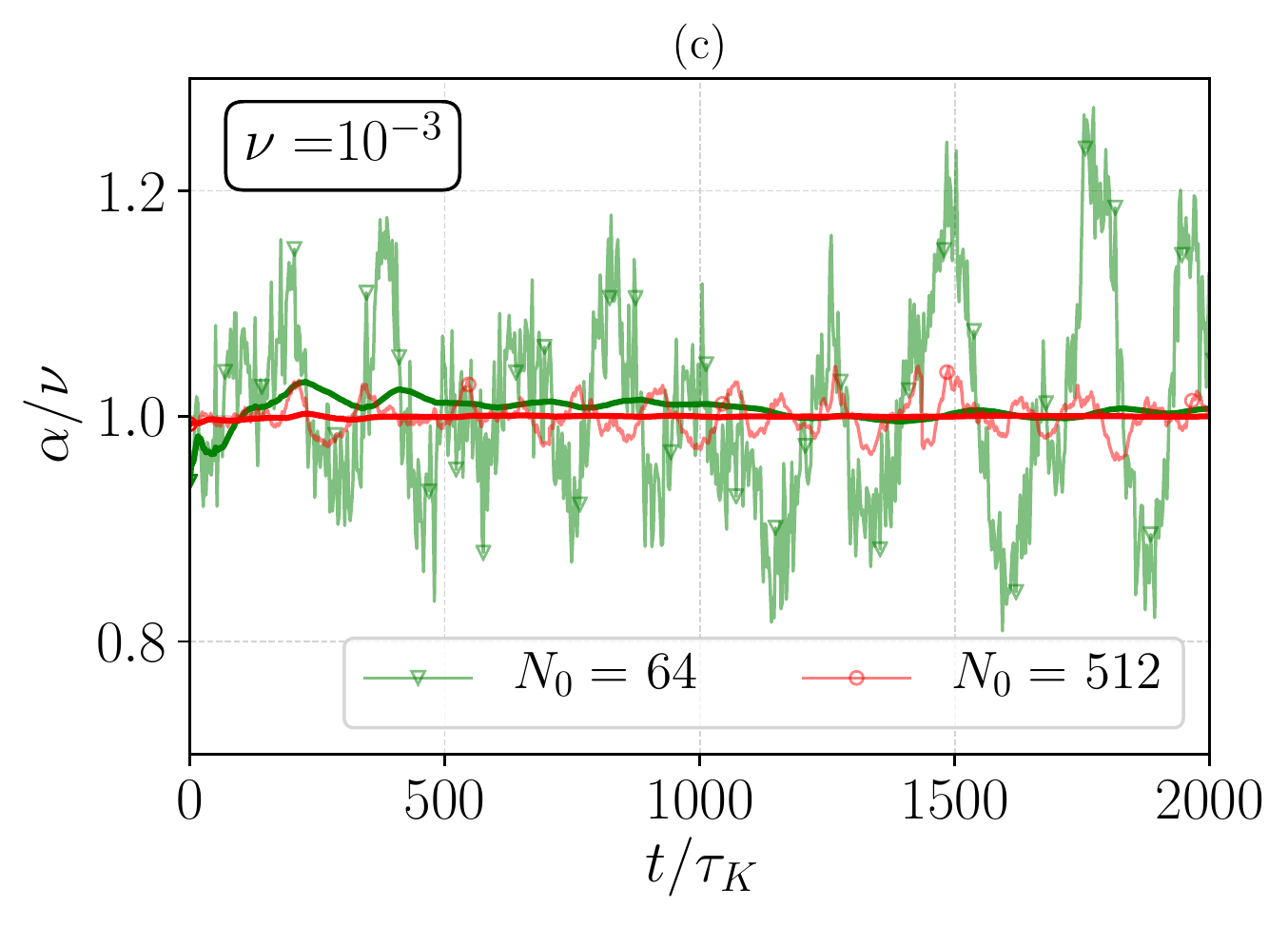}
		\includegraphics[width=.43\linewidth]{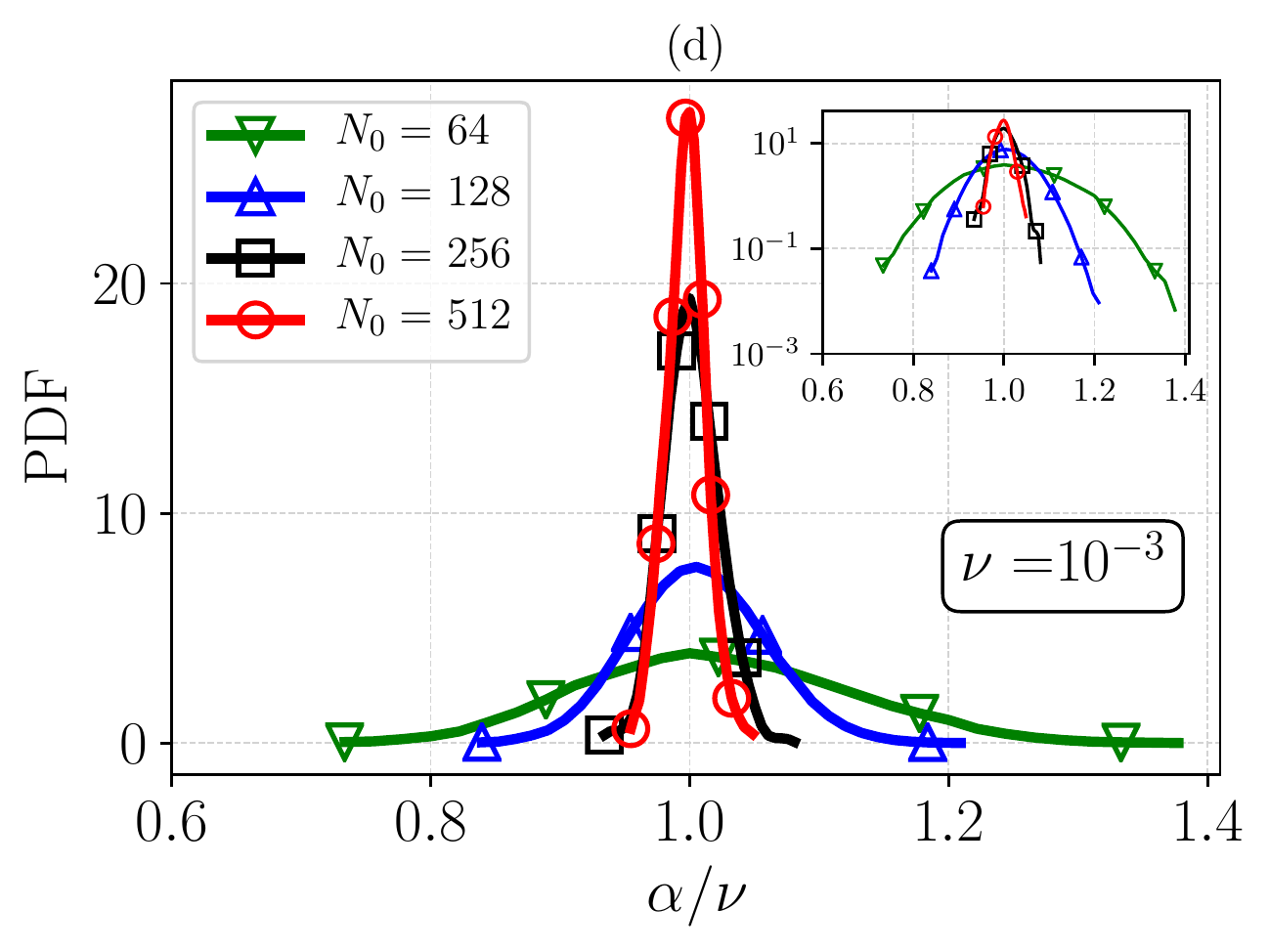}
		\caption{Statistics of $\alpha$ for $\I$. (a), (c) Temporal evolution of $\alpha/\nu$ as a function of  time, which is rescaled by the Kolmogorov time scale $\tau_K$, at (a) $\nu=10^{-2}$ and (c) $\nu=10^{-3}$. In solid line, green (for $N_0=64$) and red (for $N_0=512$) is the running average of $\alpha/\nu$. The fact that $\langle \alpha/\nu \rangle = 1$ within 0.1\% in $\I$ is remarkable (in spite of the difference in fluctuations with $\R$), as $\a$ is a non-local observable, hence not directly expected by conjecture 2. (b), (d) Probability density function of $\alpha/\nu$ obtained considering the whole available statistics for each $N_0$, at (b) $\nu=10^{-2}$ and (d) $\nu=10^{-3}$. The inset is the PDF in logarithmic scale. }
		\label{fig:alpha_fluct_INS} 
	\end{figure*}
	
	\subsection{Numerical results  about the sign of $\alpha$ in $\I$}
	\label{sec:num_res_alpha_I}
	
	Interestingly enough, $\alpha(\uu)$ is an observable that can be measured also in the  $\I$ ensemble, by computing Eq.~\eqref{e2.2}. In Fig.~\ref{fig:alpha_fluct_INS} we present the cases for two viscosities, $\nu=10^{-2}$ [(a) and (b)] and $\nu=10^{-3}$ [(c) and (d)]. In Figs.~\ref{fig:alpha_fluct_INS}(a), \ref{fig:alpha_fluct_INS}(c) we show the time series of $\alpha(\uu)/\nu$ for a time window scaled by the Kolmogorov time scale $\tau_K$, and in Figs.~\ref{fig:alpha_fluct_INS}(b), \ref{fig:alpha_fluct_INS}(d) the corresponding PDF for different different resolutions $N_0$. 
	
	First, from Figs.~\ref{fig:alpha_fluct_INS}(a)--\ref{fig:alpha_fluct_INS}(c)  we note that also in $\I$ the non-trivial result $\media{\a(\uu)}^{\I,N}_\nu = \nu$ holds (see also Sec.~\ref{sec:mean_alpha}). On the other hand, from panels Figs.~\ref{fig:alpha_fluct_INS}(b)--\ref{fig:alpha_fluct_INS}(d) we observe some  different behavior from  the corresponding PDF of the $\R$ ensemble in particular for the case of $\nu=10^{-3}$. In the latter case the  high wave number range starts to depart from the hydrodynamic regimes and probably has a direct impact on the fluctuations of $\alpha$.  
	
	\section{Conclusions}\label{sec:Conclusions}
	
	We present a detailed numerical study of the 3D reversible Navier-Stokes equation [see Eq.~\eqref{e2.2}] obtained by imposing the conservation of enstrophy $D$ via a thermostat, and compare our results with the output obtained from the corresponding [see Eq.~\eqref{e1.4}] standard irreversible NSE case. We investigate two conjectures concerning the equivalence of the two ensembles. These conjectures, which had been previously presented in \cite{Ga020b,Ga021a}, pertain to two limiting conditions. The first equivalence is studied by keeping the flow evolving on a finite set of modes, and by decreasing the viscosity of the irreversible system or, equivalently, increasing the total enstrophy of the reversible one. In this limit, which leads to the regime that we have dubbed quasithermalized in the sense of \cite{AB020}, the equivalence between the two ensembles is more and more accurate with decreasing $\nu$  for any fixed $N$ and for all scales, thereby confirming conjecture 1. 
	
	 We emphasize that conjecture 2, which is the most relevant to the physics of turbulence, had never been tested in 3D Navier-Stokes in such a detail and rigor, and only partially in 2D Navier-Stokes. In particular, opposed to what was presented in \cite{JC020}, here we have studied the limit of larger and larger values of $N$ while keeping a fixed value for the viscosity, and then explored the $\nu\to 0$ limit. In this regime we empirically observe that the equivalence between the two ensembles is restricted to ``local'' observables having support in a region of the Fourier space restricted by a typical wave number $K$, where $K$ is smaller than, but proportional to, $k_\nu$. The latter observation has been made possible by extending the class of observable studied in \cite{JC020} to  high order moments of the Fourier energy content and to its whole probability distribution function. 
	Our results are consistent against changes in $N$ and by decreasing the time discretization at least as far as we could test within our numerical capacities. The inertial-range equivalence between the two ensembles in the hydrodynamic limit is a further important confirmation of the robustness and universality of Navier-Stokes equations against changing of the dissipative mechanisms, at least concerning wave numbers smaller than $c_\nu k_\nu$. This is particularly non-trivial as the dissipative term of $\R$ is highly nonlinear and non-local.
	
 From our numerical results the prefactor $c_\nu$ entering in conjecture 2 is close to a  constant $\approx 1/8$ or $\approx 1$ depending on the observable (see discussion in the end of Sec.~\ref{sec:locality_k}) and independently of $N$. The question of the scaling in the $\nu \to 0$ limit remains an important open question that will require more numerical studies. 
	
Note that, in Remark \ref{item4conj2} of Sec.~\ref{sec:conjectures}, we mentioned a stronger version of conjecture 2, presented in \cite{Ga020b,Ga021a}, where $c_\nu=\infty$ for all $\nu$. Our results here make  it appear doubtful that such conjecture could hold in such a strong sense. However, a study for a different fixed observable (instead of the enstrophy $D$), of the time scale, of the larger values $N$, and of the integration precision necessary to reveal the $\nu$ dependence of $c_\n$, or even $c_\nu=\infty$ for all $\n$,  may be necessary to reach a firm conclusion.

		Furthermore, we have presented a numerical empirical study of the PDF of the fluctuating viscosity for the reversible case, showing the existence of a trend towards less and less probable negative events by increasing the Reynolds in the hydrodynamical regime, similarly to what had been observed in a reversible shell model \cite{BCDGL018}. The problem of the sign of $\a$ is related to that of the divergence of the phase-space contraction rate $\sigma$ \cite{Ga021a} and the problem of measuring the large deviations of $\sigma$ is extremely delicate due to the fact that negative events are expected to happen with extremely small probability. 
		We remark that, a recently published paper has presented results on the probability of observing negative values of $\alpha$ (see Fig. 4 of \cite{JC020}) that are in contrast with what has been reported for the hydrodynamical limit ($N \to \infty$ first) in this work. In particular, in \cite{JC020} a transition in the shape of the PDF is observed for large numerical grids showing  a non negligible probability to have negative events.

		Finally, by studying the average of $\a$ for the $\I$, we have shown that equivalence can also hold for a  non local observable, which is a non trivial application of the conjectures. So, it would be natural to test whether equivalence might be extended to other important non local observables. For instance,  the Lyapunov exponents, as done in a simpler  model in   \cite{GL014}, and in 2D NSE, \cite{Ga020b,Ga018}, where equivalence of the local Lyapunov spectra \footnote{obtained from the Jacobian matrix, formally $J(\uu) = \frac{\partial \dot\uu}{\partial \uu}$, by averaging over time the spectrum $\lambda_0(\uu)$, $\lambda_1(\uu)$, \dots,  of the symmetric part of $J(u)$.} is observed in several cases, in spite of its non local nature. Further numerical investigations by increasing both the cut-off $N$, and the runs duration  would be extremely useful to better elucidate the equivalence between the two ensembles in the asymptotic limit when $N\to \infty$ first and $\nu \to 0$ later, the so called fully developed turbulence.

	\begin{acknowledgements}
		This work was supported by the European Research Council (ERC) under the European Union’s Horizon 2020 research and innovation programme (Grant Agreement No. 882340). V.L.~acknowledges the support received from the EPSRC Project No.~EP/T018178/1 and from the EU Horizon 2020 project TiPES (Grant No. 820970). GM is grateful to Fabio Bonaccorso, Michele Buzzicotti and Mateo Lulli for useful discussions. This work has benefited from computational resources provided by the INFN-CINECA high performance computing facilities, by the NewTURB academic cluster at the University of Rome ``Tor Vergata'', and, to a smaller extent, from INFN-FARM cluster at University of Rome ``La Sapienza''.
	\end{acknowledgements}
		
		\appendix
		
		\section{Bound on $\a$} 
		\label{appendixA1}

		We recall Eq.~\equ{e1.5}, for which it is well known, that it leads to the {\it a priori} bounds
		\be
		\eqalign{
			&||\uu^N(t)||^2_2\le \max(E_0, ({F_0}\n^{-1})^2)\defi \EE\cr
			&\ig^t_0\,d\t ||\BDpr { {\uu^N}}(\t)||_2^2\le
			\left(\frac12\EE+t \sqrt{F_0 E_0}\right)\n^{-1}\cr}\Eq{e1.6}\ee
		satisfied (for all $N$) by the solution $\V u$ in terms of the square $L_2$ norms $E_0= ||\uu(0)||_2^2=\sum_{\b,\kk}|u_{\b,\kk}|^2=\frac1{(2\p)^3}\int |u(\xx)|^2d\xx$ and $F_0=||\ff||_2$, see for instance Proposition 1, Sec.3.2 in \cite{Ga002}. There is no problem about the forcing term $W(\uu)=\int \D\ff\cdot\uu \,d\xx$ in Eq.~\equ{e2.2} as Eq.~\equ{e1.6} implies, if $\G=\int (\D\uu)^2 \,d\xx\ge D$, a bound on it: $K^2 F\sqrt{\EE}\G^{-1}\le C D^{-\frac12}$.  The other term has in the numerator $\L(\uu)=-\int (\uu\cdot\BDpr \uu)\cdot\D\uu \,d\xx$ which is bounded via the H\"older inequality with exponents $4,4,2$ by
		\be |\L(\uu)|\le \left(\int ||\uu||_2^4\,d\xx\right)^{\frac14} \left(\int ||\BDpr\uu||_2^4\,d\xx\right)^{\frac14} \left(\int ||\D\uu||_2^2\,d\xx\right)^{\frac12}\Eq{e3.1}\ee
		and the three factors can be bounded via Sobolev's inequality \cite{So963,CKN982}: {\rm if} $2\le q\le6,\ a={3\over 4}(q-2)$ then
		\be\kern-5mm\eqalign{
			\ig_{B_r}||\uu||_2^q\,d\xx \le&
			C^S_q\Big[\left(\ig_{B_r}||\BDpr \uu||_2^2\,d\xx \right)^a\cdot\left(\ig_{B_r}
			||\uu||_2^2\,d\xx \right)^{{q/2}-a}\kern-5mm+\cr
			&+r^{-2a}\left(\ig_{B_r}||\uu||_2^2\,d\xx\right)^{q/2}
			\Big]\qquad \cr}\Eq{e3.2}\ee
		where $B_r$ is a sphere of radius $r$ and the integrals are performed with respect to $d\xx$. The $C^S_q$ is a suitable constant. The second term of the right hand side can be omitted if $\V u$ has zero average over $B_r$.
		
		Therefore in the above case, where $\uu,\BDpr\uu,\D\uu$ do have zero average, choose $q=4,a=\frac32$, $B_r=[0,2\p]^3$;  calling $\lis C^S_4=(C^S_4)^{\frac14}$ and $\G=\int (\D\uu)^2\,d\xx$, it is:
		\be\eqalign{
			\L(\uu)&\le \lis C^S_4 (\EE^{\frac18} D^{\frac38})\,\lis C^S_4
			( D^{\frac18} \G^{\frac38})\G^{\frac12}\le
			\lis C \EE^{\frac18} D^{\frac12} \G^{\frac78}\cr} 
		\Eq{e3.3}
		\ee
		so that 
		\be\kern-3mm\eqalign{ |\a(\uu)|\le& C_1
			(\EE^{\frac18}D^{\frac12}\G^{-\frac18}+\EE^{\frac12}\G^{-1})\cr
			\le& C(\EE^{\frac18}
			D^{\frac38}+\EE^{\frac12}D^{-1})<C_2 (\sqrt{D}+{\sqrt{D}}^{-1})
			\cr}\Eq{e3.4}
		\ee
		and $|\a(\uu(t))|$ is thus bounded, for any solution and any $N$, by a constant: $|\a|\le \k$.

		\def\SEC{Reversible Navier-Stokes global smoothness}
		\section{\SEC}
		\label{seca2}
		
		Let $\e <\a(\uu(t))\le \k $  and suppose that the initial data
		$\uu(0)$ and $\ff$ satisfy $||\uu_{\kk}||_2,||\ff_\kk||_2< c_p k^{-p}$ for all
		$p>0$ (recall that we consider only initial data and force with a finite
		number, $\le N$, of modes, for simplicity). Let $a(t,\t)=\int_\t^t
		\a(\uu(t')dt'$; then $\e (t-\t)<a(t,\t)<\k (t-\t)$. 
		
		Following, for instance \cite{Ga002}, we can write $\uu_\kk(t)=e^{-a(t,0) k^2}\uu_\kk(0)+\int_0^t e^{-a(t,\t)k^2}( {\rm N}_\kk(\uu(\t))+\ff_\kk)d\t$, where ${\rm N}_\kk(\uu(\t))$ is the non-linear term of the NSE.
		Therefore, the sum of the first and last term can be bounded by $\frac{2c_p
		}{k^p}$ while the integral is bounded by
		\be \eqalign{& \int_0^t e^{-\e
				k^2(t-\t)}\sum_{\pp+\qq=\kk} ||\uu_{\pp}\cdot
			\qq||_2 ||\uu_{\qq}||_2 \,d\t \cr& \le \frac{\sqrt\EE
				\sqrt D(1-e^{-\e k^2 t})}{\e k^2}\cr} \Eq{a2.1}\ee
		so that adding the two bounds: $||\uu_\kk||_2^2< \frac{C_2}{k^2}$ for a
		suitable $C_2$. Therefore, again, $||\uu_\kk(t)||_2$ can be bounded
		by adding $\frac{2c_p }{k^p}$ and a bound on

		\be \eqalign{&\int_0^t e^{-\e k^2(t-\t)}
			\sum_{\pp+\qq=\kk}
			\frac{{||\pp||_2}||\uu_{\pp}||_2\, ||\qq||_2^2||\uu_{\qq}||_2}{||\pp||_2||\qq||_2} \,d\t
			\cr}\Eq{a2.2}
		\ee
		%
		\hglue.2cm A bound on the latter integral is obtained via the Schwartz inequality and
		the remark that 
		$\pp+\qq=\kk$ implies $||\pp||_2||\qq||_2\ge \frac{k_0}2 ||\kk||_2, k_0=1$, and
		\be \kern-5mm\eqalign{
			&\sum_{\pp+\qq=\kk}\kern-3mm
			\frac{{||\pp||_2}||\uu_{\pp}||_2 ||\qq||_2^2||\uu_{\qq}||_2}{||\pp||_2||\qq||_2}
			\le \frac{2 C_2}{k_0} \sum_{\pp+\qq=\kk}\kern-3mm
			\frac{||\pp||_2||\uu_{\pp}||_2}{||\pp||_2||\qq||_2}\cr
			&\le \frac{2 C_2}{k_0} \sqrt{D} \left(\sum_{\pp+\qq=\kk}
			\frac1{(||\pp||_2||\qq||_2)^2}\right)^{\frac12}
			\cr
			&\le \left(\frac{2 C_2}{k_0}\right)^{1+\frac18} ||\kk||_2^{-\frac14} \sqrt{D}
			\left(\sum_{\pp+\qq=\kk}\frac1{(||\pp||_2||\qq||_2)^{2-\frac14}}\right)^{\frac12}
			\cr
			&\le \left(\frac{2 C_2}{k_0}\right)^{1+\frac18} \sqrt{D}||\kk||_2^{-\frac14} 
			\left(\sum_\nn \frac1{||\nn||_2^{4-\frac12}}\right)^{\frac12}\cr}\Eq{a2.3}\ee
		where $\pp$ has been changed to $\nn$ just to make clear that summing
		over $\pp+\qq=\kk$ allows using the Schwartz inequality. Hence
		integration over $t$, as in Eq.~\equ{a2.1}, yields
		\be||\uu_\kk(t)||_2 \le \frac{\g_1}{k^{2+\frac14}}\Eq{a2.4}\ee
		\hglue.2cm Thus if $D$ is finite the bound $||\uu_\kk||_2<\g k^{-2}$, Eq.~\equ{a2.1},
		can be improved into $||\uu_\kk||_2<\g_1 k^{-2-\frac14}$. Iterating a {\it
			autoregularization} phenomenon sets in and
		\be ||\uu_\kk(t)||_2 \le \frac{\g_p}{k^{2+\frac14 p}}\qquad {\rm for\ all}
		\ p\ge1\Eq{a2.5}\ee
		so that $\uu(t)$ is a $C^\infty$-functions and all its derivatives can be
		bounded in terms of the enstrophy $D$, uniformly in $N$. See Sec. 3.3 in
		\cite{Ga002} for related results on the classic autoregularization.

		\begin{table}
			\caption{ Values of various observables corresponding to the numerical simulations, labeled as R\# indicating the run numbering, with different kinematic viscosities and cut-offs. All resulting parameters are the same up to statistical errors (which are included inside the parentheses for the error of the last digit) for both $\I$ and  $\R$, except for those cases separated as $\R|\I$. $D/\DD$ is an abbreviation for $D/\DD \equiv D / \media{\DD(\uu)}_\n^{\I,N}$. The symbols next to the Run numbering correspond to the statistical regimes; $\bigcirc$: Hydrodynamic, $\triangle$: Crossover, $\square$: quasithermalized. }
			\centering
			\begin{tabular}{c c c c c c}
				\toprule[1pt]
				$N_0=64$  & R$1\bigcirc$ \qquad & R$2\bigcirc$ \qquad & R$3\triangle$ \qquad & R$4\square$ \qquad & R$5\square$ \qquad\\
				$\nu$ & $5 \times 10^{-2}$ &    $10^{-2}$  & $10^{-3}$ &    $10^{-4}$  & $10^{-5}$\\\midrule[0.5pt]
				$\langle\alpha\rangle/\nu$ & 1.015(7) &    0.999(3)  & 1.000(4) &    1.000(8)  & 0.93(20)\\
				$D/\DD$ & 0.998(2)&  0.999(1)&  0.999(1)&  1.000(1)&  1.01(2) \\
				$u_{\rm rms}$ &  1.22(1) & 1.30(1)    & 1.46(1) & 2.92(1) & 7.00(1)\\		
				Re  & 43   & 185 & 1705 & $7.9 \times 10^{3}$ & $1.2 \times 10^{5}$\\
				Re$_{\lambda}$ & 30   & 77 & 300 &-- &--\\
				$\varepsilon$ & 0.74 & 0.72  & 0.76 & 0.66 & 0.39\\
				$k_\nu$ &   8   & 29  & 165  & -- & -- \\
				$\lambda$ & 1.23 & 0.59 & 0.21 & --& --\\
				$\ell$ & 1.77 & 1.43 & 1.16 & 0.27 & 0.17 \\
				$T/T_\ell$ & 8300  & $1.3 \times 10^{4}$ & $2.3 \times 10^{4}$ &$1.8 \times 10^{5}$ & $5.8 \times 10^{5}$\\
			\end{tabular}
			
			\vspace{0.1cm}
			
			\begin{tabular}{c c c c c c}
				\toprule[1pt]
				$N_0=128$  & R$6\bigcirc$ \qquad & R$7\bigcirc$ \qquad & R$8\triangle$ \qquad & R$9\triangle$ \qquad & R$10\square$ \qquad\\
				$\nu$ & $5 \times 10^{-2}$ &    $10^{-2}$  & $10^{-3}$ &    $10^{-4}$  & $10^{-5}$\\ \midrule[0.5pt]
				$\langle\alpha\rangle/\nu$ & 1.01(3)& 0.995(5)& 1.011(6)& 0.99(1)& 0.99(2)\\
				$D/\DD$ & 1.008(2)&  0.998(1)&  0.996(2)&  0.999(2)&  1.027(1) \\
				$u_{\rm rms}$ &  1.22(4)& 1.29(3)& 1.36(3)& 1.88(4)& 4.61(1)\\		
				Re  & 43   & 183 & 1850 & $1.4 \times 10^{4}$ & $6.1 \times 10^{4}$\\
				Re$_{\lambda}$ & 30& 76& 267& 1570& --\\
				$\varepsilon$ & 0.75& 0.71& 0.73& 0.76& 0.68\\
				$k_\nu$ &   8   & 29  & 164  & 932 & -- \\
				$\lambda$ & 1.22& 0.59& 0.20& 0.08& --\\
				$\ell$ & 1.78 & 1.42 & 1.36 & 0.72 & 0.13 \\
				$T/T_\ell$ & 760  & 4690 & 4950 &$10^{4}$ & $1.1 \times 10^{5}$\\
			\end{tabular}
			\vspace{0.1cm}
			
			\begin{tabular}{c c c c c c}
				\toprule[1pt]
				$N_0=256$  & R$11\bigcirc$ \qquad & R$12\bigcirc$ \qquad & R$13\triangle$ \qquad & R$14\triangle$ \qquad & R$15\square$ \qquad\\
				$\nu$ & $5 \times 10^{-2}$ &    $10^{-2}$  & $10^{-3}$ &    $10^{-4}$  & $10^{-5}$\\ \midrule[0.5pt]
				$\langle\alpha\rangle/\nu$ & 1.06(4)& 0.99(2)& 1.00(2)& 1.00(2)& 0.94(2)\\
				$D/\DD$ & 1.008(2)&  1.008(1)&  1.008(2)&  1.019(2)&  1.034(2) \\
				$u_{\rm rms}$ &  1.25$|$1.21 & 1.29(5)& 1.35(5)& 1.46(5)& 2.70(6)\\		
				Re  & 44   & 184 & 1800 & $1.7 \times 10^{4}$ & $ 10^{5}$\\
				Re$_{\lambda}$ & 31& 76& 260& 973& --\\
				$\varepsilon$ & 0.78$|$0.73& 0.71& 0.73& 0.71& 0.74$|$0.79\\
				$k_\nu$ &   8   & 29  & 164  & 919 & -- \\
				$\lambda$ & 1.26$|$1.22& 0.59& 0.19& 0.06& --\\
				$\ell$ & 1.80& 1.42& 1.33& 1.18& 0.36 \\
				$T/T_\ell$ & 374  & 680& 760& 800& 3950\\
			\end{tabular}
			
			\vspace{0.1cm}
			
			\begin{tabular}{c c c c c c c}
				\toprule[1pt]
				R$16\bigcirc$ &	 $\nu$ & $\langle\alpha\rangle/\nu$ & $D/\DD$ & $u_{\rm rms}$ & Re & Re$_{\lambda}$\\
				\midrule[0.5pt]
				&	 $10^{-3}$ &    0.97(3)  & 0.997(2)& 1.35(5) &    1800  & 260 \vspace{0.1cm}\\ 
				$N_0=512$ &	 $\varepsilon$ & $k_\nu$& $\lambda$& $\ell$& $T/T_\ell$ & 	\\ \midrule[0.5pt]
				&	 0.71(7) &   163  & 0.19 &    1.35  & 400 &\\
				\bottomrule[0.5pt]
			\end{tabular}
			\label{tab:parameters}
		\end{table}

		\section{Numerical parameters}\label{appendixTable}

		Table \ref{tab:parameters} summarizes the values of various observables corresponding to the runs with different kinematic viscosities and cut-offs. The parentheses present the statistical error of the last digit in a given value.  We define the following: physical length of the container $L=2\pi$, root-mean-square velocity $u_{\rm rms}= \sqrt{ \langle ||\uu||_2^2 \rangle/3}$,
		averaged rate of energy dissipation
		$\varepsilon= \nu \left\langle D\right\rangle$,
		$k_\nu = \left(\frac{\varepsilon}{\nu^3}\right)^{1/4}$ is the Kolmogorov scale,
		Taylor length $\lambda = u_{\rm rms} \sqrt{\frac{15}{\langle D\rangle}}$, integral length $\ell=\frac{3L}{8} \langle\sum_\kk\, k^{-1}E(\kk)/ \sum_\kk\, E(\kk)\rangle$ (see Eq.~\ref{eq:E_k}),
		Taylor-Reynolds number $\textrm{Re}_{\lambda}=
		\frac{u_{\rm rms} \lambda}{\nu}$, Reynolds number $\textrm{Re}= \frac{u_{\rm rms} \ell}{\nu}$, large-eddy turnover time
		$T_\ell=\frac{\ell}{u_{\rm rms}}$.  
		
		We present the final $D/\DD$ ratio, which is an abbreviation for $D/\DD \equiv D / \media{\DD(\uu)}_\n^{\I,N}$. Ideally, this should be 1, and it is indeed unity when we start the $\R$ simulation based on the ensemble average $\media{\DD(\uu)}_\n^{\I,N}$ from $\I$. But for almost all runs, we increased the statistics of $\I$, which eventually improved $\media{\DD(\uu)}_\n^{\I,N}$. Therefore the $D/\DD$ ratio gives an estimate of the quality of each collection  $\EE^{\R,N}$ of $\R$ runs with respect to the collection  $\EE^{\I,N}$ for a particular $\nu$. A significant $D/\DD$ discrepancy would also explain possible discrepancies in $\langle\alpha\rangle/\nu$ and accordingly other ratios of observables.

		The ratio $T/T_\ell$, where $T$ is the total length of the simulation in physical units, is essentially an indicator of the statistical significance of the generated ensemble. Empty entries (filled with a ``--'') are met in cases at the quasithermalized regime for particular observables of which the definitions are relevant in the hydrodynamic regime (e.g. $k_\nu, \, \lambda$).

		An interesting fact about Table~\ref{tab:parameters} is that the averaged values for each entry turn out to be the same on average, up to statistical errors, for both $\R$ and $\I$. Showing one value, implies that is it the same for $\R$ and $\I$ at the particular $N_0$ and $\nu$. When the averages are not the same within statistical errors we show both estimates in the form $\R|\I$. This happened in the case of $N_0=256$ in two cases, namely R$11\bigcirc$ and R$15\square$, see e.g. $\varepsilon$. For R$11\bigcirc$ we attribute this to poor statistics, notice $T/T_\ell = 374$, which is the lowest, and for R$15\square$ we attribute to a poor estimation of $\media{\DD(\uu)}_\n^{\I,N}$, notice $D/\DD=1.034(2)$, which is the highest.

		\bibliography{Bibliography}

\begin{thebibliography}{41}%
\makeatletter
\providecommand \@ifxundefined [1]{%
 \@ifx{#1\undefined}
}%
\providecommand \@ifnum [1]{%
 \ifnum #1\expandafter \@firstoftwo
 \else \expandafter \@secondoftwo
 \fi
}%
\providecommand \@ifx [1]{%
 \ifx #1\expandafter \@firstoftwo
 \else \expandafter \@secondoftwo
 \fi
}%
\providecommand \natexlab [1]{#1}%
\providecommand \enquote  [1]{``#1''}%
\providecommand \bibnamefont  [1]{#1}%
\providecommand \bibfnamefont [1]{#1}%
\providecommand \citenamefont [1]{#1}%
\providecommand \href@noop [0]{\@secondoftwo}%
\providecommand \href [0]{\begingroup \@sanitize@url \@href}%
\providecommand \@href[1]{\@@startlink{#1}\@@href}%
\providecommand \@@href[1]{\endgroup#1\@@endlink}%
\providecommand \@sanitize@url [0]{\catcode `\\12\catcode `\$12\catcode
  `\&12\catcode `\#12\catcode `\^12\catcode `\_12\catcode `\%12\relax}%
\providecommand \@@startlink[1]{}%
\providecommand \@@endlink[0]{}%
\providecommand \url  [0]{\begingroup\@sanitize@url \@url }%
\providecommand \@url [1]{\endgroup\@href {#1}{\urlprefix }}%
\providecommand \urlprefix  [0]{URL }%
\providecommand \Eprint [0]{\href }%
\providecommand \doibase [0]{http://dx.doi.org/}%
\providecommand \selectlanguage [0]{\@gobble}%
\providecommand \bibinfo  [0]{\@secondoftwo}%
\providecommand \bibfield  [0]{\@secondoftwo}%
\providecommand \translation [1]{[#1]}%
\providecommand \BibitemOpen [0]{}%
\providecommand \bibitemStop [0]{}%
\providecommand \bibitemNoStop [0]{.\EOS\space}%
\providecommand \EOS [0]{\spacefactor3000\relax}%
\providecommand \BibitemShut  [1]{\csname bibitem#1\endcsname}%
\let\auto@bib@innerbib\@empty
\bibitem [{\citenamefont {Gallavotti}(2014)}]{Ga013b}%
  \BibitemOpen
  \bibfield  {author} {\bibinfo {author} {\bibfnamefont {G.}~\bibnamefont
  {Gallavotti}},\ }\href@noop {} {\emph {\bibinfo {title} {Nonequilibrium and
  irreversibility}}},\ Theoretical and Mathematical Physics\ (\bibinfo
  {publisher} {Springer-Verlag},\ \bibinfo {year} {2014})\BibitemShut {NoStop}%
\bibitem [{\citenamefont {Gallavotti}(2020{\natexlab{a}})}]{Ga020b}%
  \BibitemOpen
  \bibfield  {author} {\bibinfo {author} {\bibfnamefont {G.}~\bibnamefont
  {Gallavotti}},\ }\href {\doibase 10.1007/s10955-019-02376-3} {\bibfield
  {journal} {\bibinfo  {journal} {Journal of Statistical Physics}\ }\textbf
  {\bibinfo {volume} {180}},\ \bibinfo {pages} {172} (\bibinfo {year}
  {2020}{\natexlab{a}})}\BibitemShut {NoStop}%
\bibitem [{\citenamefont {Frisch}(1995)}]{Frisch1995}%
  \BibitemOpen
  \bibfield  {author} {\bibinfo {author} {\bibfnamefont {U.}~\bibnamefont
  {Frisch}},\ }\href {\doibase 10.1017/CBO9781139170666} {\emph {\bibinfo
  {title} {Turbulence: The Legacy of A. N. Kolmogorov}}}\ (\bibinfo
  {publisher} {Cambridge University Press},\ \bibinfo {year}
  {1995})\BibitemShut {NoStop}%
\bibitem [{\citenamefont {Pope}(2000)}]{Pope2000}%
  \BibitemOpen
  \bibfield  {author} {\bibinfo {author} {\bibfnamefont {S.~B.}\ \bibnamefont
  {Pope}},\ }\href {\doibase 10.1017/CBO9780511840531} {\emph {\bibinfo {title}
  {Turbulent Flows}}}\ (\bibinfo  {publisher} {Cambridge University Press},\
  \bibinfo {year} {2000})\BibitemShut {NoStop}%
\bibitem [{\citenamefont {Salmon}(1998)}]{Salmon1998}%
  \BibitemOpen
  \bibfield  {author} {\bibinfo {author} {\bibfnamefont {R.}~\bibnamefont
  {Salmon}},\ }\href@noop {} {\emph {\bibinfo {title} {Lectures on geophysical
  fluid dynamics}}}\ (\bibinfo  {publisher} {Oxford University Press},\
  \bibinfo {year} {1998})\BibitemShut {NoStop}%
\bibitem [{\citenamefont {Alexakis}\ and\ \citenamefont
  {Biferale}(2018)}]{alexakis2018cascades}%
  \BibitemOpen
  \bibfield  {author} {\bibinfo {author} {\bibfnamefont {A.}~\bibnamefont
  {Alexakis}}\ and\ \bibinfo {author} {\bibfnamefont {L.}~\bibnamefont
  {Biferale}},\ }\href@noop {} {\bibfield  {journal} {\bibinfo  {journal}
  {Physics Reports}\ }\textbf {\bibinfo {volume} {767}},\ \bibinfo {pages} {1}
  (\bibinfo {year} {2018})}\BibitemShut {NoStop}%
\bibitem [{\citenamefont {Frisch}\ \emph {et~al.}(2008)\citenamefont {Frisch},
  \citenamefont {Kurien}, \citenamefont {Pandit}, \citenamefont {Pauls},
  \citenamefont {Ray}, \citenamefont {Wirth},\ and\ \citenamefont
  {Zhu}}]{Frisch2008}%
  \BibitemOpen
  \bibfield  {author} {\bibinfo {author} {\bibfnamefont {U.}~\bibnamefont
  {Frisch}}, \bibinfo {author} {\bibfnamefont {S.}~\bibnamefont {Kurien}},
  \bibinfo {author} {\bibfnamefont {R.}~\bibnamefont {Pandit}}, \bibinfo
  {author} {\bibfnamefont {W.}~\bibnamefont {Pauls}}, \bibinfo {author}
  {\bibfnamefont {S.~S.}\ \bibnamefont {Ray}}, \bibinfo {author} {\bibfnamefont
  {A.}~\bibnamefont {Wirth}}, \ and\ \bibinfo {author} {\bibfnamefont {J.-Z.}\
  \bibnamefont {Zhu}},\ }\href {\doibase 10.1103/PhysRevLett.101.144501}
  {\bibfield  {journal} {\bibinfo  {journal} {Phys. Rev. Lett.}\ }\textbf
  {\bibinfo {volume} {101}},\ \bibinfo {pages} {144501} (\bibinfo {year}
  {2008})}\BibitemShut {NoStop}%
\bibitem [{\citenamefont {Cercignani}(1988)}]{Cercignani1988}%
  \BibitemOpen
  \bibfield  {author} {\bibinfo {author} {\bibfnamefont {C.}~\bibnamefont
  {Cercignani}},\ }\enquote {\bibinfo {title} {The boltzmann equation},}\ in\
  \href {\doibase 10.1007/978-1-4612-1039-9_2} {\emph {\bibinfo {booktitle}
  {The Boltzmann Equation and Its Applications}}}\ (\bibinfo  {publisher}
  {Springer New York},\ \bibinfo {address} {New York, NY},\ \bibinfo {year}
  {1988})\ pp.\ \bibinfo {pages} {40--103}\BibitemShut {NoStop}%
\bibitem [{\citenamefont {Zwanzig}(2001)}]{Zwanzig2001}%
  \BibitemOpen
  \bibfield  {author} {\bibinfo {author} {\bibfnamefont {R.}~\bibnamefont
  {Zwanzig}},\ }\href@noop {} {\emph {\bibinfo {title} {Nonequilibrium
  statistical mechanics}}}\ (\bibinfo  {publisher} {Oxford university press},\
  \bibinfo {year} {2001})\BibitemShut {NoStop}%
\bibitem [{\citenamefont {Onsager}\ and\ \citenamefont
  {Machlup}(1953)}]{OM953a}%
  \BibitemOpen
  \bibfield  {author} {\bibinfo {author} {\bibfnamefont {L.}~\bibnamefont
  {Onsager}}\ and\ \bibinfo {author} {\bibfnamefont {S.}~\bibnamefont
  {Machlup}},\ }\href {\doibase 10.1103/PhysRev.91.1505} {\bibfield  {journal}
  {\bibinfo  {journal} {Phys. Rev.}\ }\textbf {\bibinfo {volume} {91}},\
  \bibinfo {pages} {1505} (\bibinfo {year} {1953})}\BibitemShut {NoStop}%
\bibitem [{\citenamefont {Bertini}\ \emph {et~al.}(2002)\citenamefont
  {Bertini}, \citenamefont {De~Sole}, \citenamefont {Gabrielli}, \citenamefont
  {Jona-Lasinio},\ and\ \citenamefont {Landim}}]{Bertini2002}%
  \BibitemOpen
  \bibfield  {author} {\bibinfo {author} {\bibfnamefont {L.}~\bibnamefont
  {Bertini}}, \bibinfo {author} {\bibfnamefont {A.}~\bibnamefont {De~Sole}},
  \bibinfo {author} {\bibfnamefont {D.}~\bibnamefont {Gabrielli}}, \bibinfo
  {author} {\bibfnamefont {G.}~\bibnamefont {Jona-Lasinio}}, \ and\ \bibinfo
  {author} {\bibfnamefont {C.}~\bibnamefont {Landim}},\ }\href@noop {}
  {\bibfield  {journal} {\bibinfo  {journal} {Journal of Statistical Physics}\
  }\textbf {\bibinfo {volume} {107}},\ \bibinfo {pages} {635} (\bibinfo {year}
  {2002})}\BibitemShut {NoStop}%
\bibitem [{\citenamefont {Ruelle}(1999)}]{Ru1999}%
  \BibitemOpen
  \bibfield  {author} {\bibinfo {author} {\bibfnamefont {D.}~\bibnamefont
  {Ruelle}},\ }\href {\doibase 10.1142/4090} {\emph {\bibinfo {title}
  {Statistical Mechanics}}}\ (\bibinfo  {publisher} {CO-PUBLISHED WITH IMPERIAL
  COLLEGE PRESS},\ \bibinfo {year} {1999})\ \Eprint
  {http://arxiv.org/abs/https://www.worldscientific.com/doi/pdf/10.1142/4090}
  {https://www.worldscientific.com/doi/pdf/10.1142/4090} \BibitemShut {NoStop}%
\bibitem [{\citenamefont {Touchette}(2015)}]{Touchette2015}%
  \BibitemOpen
  \bibfield  {author} {\bibinfo {author} {\bibfnamefont {H.}~\bibnamefont
  {Touchette}},\ }\href {\doibase https://doi.org/10.1007/s10955-015-1212-2}
  {\bibfield  {journal} {\bibinfo  {journal} {Journal of Statistical Physics}\
  }\textbf {\bibinfo {volume} {159}},\ \bibinfo {pages} {987} (\bibinfo {year}
  {2015})}\BibitemShut {NoStop}%
\bibitem [{\citenamefont {Evans}\ and\ \citenamefont {Morriss}(1990)}]{EM990}%
  \BibitemOpen
  \bibfield  {author} {\bibinfo {author} {\bibfnamefont {D.~J.}\ \bibnamefont
  {Evans}}\ and\ \bibinfo {author} {\bibfnamefont {G.~P.}\ \bibnamefont
  {Morriss}},\ }\href@noop {} {\emph {\bibinfo {title} {Statistical Mechanics
  of Non{\-}equilibrium Fluids}}}\ (\bibinfo  {publisher} {Academic Press},\
  \bibinfo {address} {New-York},\ \bibinfo {year} {1990})\BibitemShut {NoStop}%
\bibitem [{\citenamefont {She}\ and\ \citenamefont {Jackson}(1993)}]{SJ993}%
  \BibitemOpen
  \bibfield  {author} {\bibinfo {author} {\bibfnamefont {Z.}~\bibnamefont
  {She}}\ and\ \bibinfo {author} {\bibfnamefont {E.}~\bibnamefont {Jackson}},\
  }\href@noop {} {\bibfield  {journal} {\bibinfo  {journal} {Physical Review
  Letters}\ }\textbf {\bibinfo {volume} {70}},\ \bibinfo {pages} {1255}
  (\bibinfo {year} {1993})}\BibitemShut {NoStop}%
\bibitem [{\citenamefont {Gallavotti}(1996)}]{Ga996b}%
  \BibitemOpen
  \bibfield  {author} {\bibinfo {author} {\bibfnamefont {G.}~\bibnamefont
  {Gallavotti}},\ }\href@noop {} {\bibfield  {journal} {\bibinfo  {journal}
  {Physics Letters A}\ }\textbf {\bibinfo {volume} {223}},\ \bibinfo {pages}
  {91} (\bibinfo {year} {1996})}\BibitemShut {NoStop}%
\bibitem [{\citenamefont {Gallavotti}(1997)}]{Ga997b}%
  \BibitemOpen
  \bibfield  {author} {\bibinfo {author} {\bibfnamefont {G.}~\bibnamefont
  {Gallavotti}},\ }\href@noop {} {\bibfield  {journal} {\bibinfo  {journal}
  {Physica D}\ }\textbf {\bibinfo {volume} {105}},\ \bibinfo {pages} {163}
  (\bibinfo {year} {1997})}\BibitemShut {NoStop}%
\bibitem [{\citenamefont {Gallavotti}\ \emph {et~al.}(2004)\citenamefont
  {Gallavotti}, \citenamefont {Rondoni},\ and\ \citenamefont {Segre}}]{GRS004}%
  \BibitemOpen
  \bibfield  {author} {\bibinfo {author} {\bibfnamefont {G.}~\bibnamefont
  {Gallavotti}}, \bibinfo {author} {\bibfnamefont {L.}~\bibnamefont {Rondoni}},
  \ and\ \bibinfo {author} {\bibfnamefont {E.}~\bibnamefont {Segre}},\
  }\href@noop {} {\bibfield  {journal} {\bibinfo  {journal} {Physica D}\
  }\textbf {\bibinfo {volume} {187}},\ \bibinfo {pages} {358} (\bibinfo {year}
  {2004})}\BibitemShut {NoStop}%
\bibitem [{\citenamefont {Gallavotti}(2020{\natexlab{b}})}]{Ga020a}%
  \BibitemOpen
  \bibfield  {author} {\bibinfo {author} {\bibfnamefont {G.}~\bibnamefont
  {Gallavotti}},\ }\href {\doibase 10.1140/epje/i2020-11961-0} {\bibfield
  {journal} {\bibinfo  {journal} {{European Physics Journal, E}}\ }\textbf
  {\bibinfo {volume} {43:37}} (\bibinfo {year} {2020}{\natexlab{b}}),\
  10.1140/epje/i2020-11961-0}\BibitemShut {NoStop}%
\bibitem [{\citenamefont {Gallavotti}\ and\ \citenamefont
  {Lucarini}(2014)}]{GL014}%
  \BibitemOpen
  \bibfield  {author} {\bibinfo {author} {\bibfnamefont {G.}~\bibnamefont
  {Gallavotti}}\ and\ \bibinfo {author} {\bibfnamefont {V.}~\bibnamefont
  {Lucarini}},\ }\href {\doibase 10.1007/s10955-014-1051-6} {\bibfield
  {journal} {\bibinfo  {journal} {Journal of Statistical Physics}\ }\textbf
  {\bibinfo {volume} {156}},\ \bibinfo {pages} {1027} (\bibinfo {year}
  {2014})}\BibitemShut {NoStop}%
\bibitem [{\citenamefont {Biferale}\ \emph {et~al.}(2018)\citenamefont
  {Biferale}, \citenamefont {Cencini}, \citenamefont {Pietro}, \citenamefont
  {Gallavotti},\ and\ \citenamefont {Lucarini}}]{BCDGL018}%
  \BibitemOpen
  \bibfield  {author} {\bibinfo {author} {\bibfnamefont {L.}~\bibnamefont
  {Biferale}}, \bibinfo {author} {\bibfnamefont {M.}~\bibnamefont {Cencini}},
  \bibinfo {author} {\bibfnamefont {M.~D.}\ \bibnamefont {Pietro}}, \bibinfo
  {author} {\bibfnamefont {G.}~\bibnamefont {Gallavotti}}, \ and\ \bibinfo
  {author} {\bibfnamefont {V.}~\bibnamefont {Lucarini}},\ }\href {\doibase
  10.1103/PhysRevE.98.012202} {\bibfield  {journal} {\bibinfo  {journal}
  {Physical Review E}\ }\textbf {\bibinfo {volume} {98}},\ \bibinfo {pages}
  {012201} (\bibinfo {year} {2018})}\BibitemShut {NoStop}%
\bibitem [{\citenamefont {Pietro}\ \emph {et~al.}(2018)\citenamefont {Pietro},
  \citenamefont {Biferale}, \citenamefont {Boffetta},\ and\ \citenamefont
  {Cencini}}]{DBBC018}%
  \BibitemOpen
  \bibfield  {author} {\bibinfo {author} {\bibfnamefont {M.~D.}\ \bibnamefont
  {Pietro}}, \bibinfo {author} {\bibfnamefont {L.}~\bibnamefont {Biferale}},
  \bibinfo {author} {\bibfnamefont {G.}~\bibnamefont {Boffetta}}, \ and\
  \bibinfo {author} {\bibfnamefont {M.}~\bibnamefont {Cencini}},\ }\href
  {\doibase 10.1140/epje/i2018-11655-2} {\bibfield  {journal} {\bibinfo
  {journal} {The European Physical Journal E}\ }\textbf {\bibinfo {volume}
  {41}},\ \bibinfo {pages} {48} (\bibinfo {year} {2018})}\BibitemShut {NoStop}%
\bibitem [{\citenamefont {Seshasayanan}\ \emph {et~al.}(2021)\citenamefont
  {Seshasayanan}, \citenamefont {Eswaran}, \citenamefont {Maji}, \citenamefont
  {Ghosh},\ and\ \citenamefont {Shukla}}]{SSMGS21}%
  \BibitemOpen
  \bibfield  {author} {\bibinfo {author} {\bibfnamefont {K.}~\bibnamefont
  {Seshasayanan}}, \bibinfo {author} {\bibfnamefont {K.~S.}\ \bibnamefont
  {Eswaran}}, \bibinfo {author} {\bibfnamefont {M.}~\bibnamefont {Maji}},
  \bibinfo {author} {\bibfnamefont {S.}~\bibnamefont {Ghosh}}, \ and\ \bibinfo
  {author} {\bibfnamefont {V.}~\bibnamefont {Shukla}},\ }\href@noop {}
  {\enquote {\bibinfo {title} {Equivalence of nonequilibrium ensembles:
  Two-dimensional turbulence with a dual cascade},}\ } (\bibinfo {year}
  {2021}),\ \Eprint {http://arxiv.org/abs/2112.12215} {arXiv:2112.12215
  [physics.flu-dyn]} \BibitemShut {NoStop}%
\bibitem [{\citenamefont {Gallavotti}(2018)}]{Ga018}%
  \BibitemOpen
  \bibfield  {author} {\bibinfo {author} {\bibfnamefont {G.}~\bibnamefont
  {Gallavotti}},\ }\href {\doibase 10.1140/epjst/e2018-700096-x} {\bibfield
  {journal} {\bibinfo  {journal} {European Physics Journal Special Topics}\
  }\textbf {\bibinfo {volume} {227}},\ \bibinfo {pages} {217} (\bibinfo {year}
  {2018})}\BibitemShut {NoStop}%
\bibitem [{\citenamefont {Gallavotti}(2019)}]{Ga019c}%
  \BibitemOpen
  \bibfield  {author} {\bibinfo {author} {\bibfnamefont {G.}~\bibnamefont
  {Gallavotti}},\ }\href {\doibase
  https://doi.org/10.1007/978-3-030-15096-9_21} {\bibfield  {journal} {\bibinfo
   {journal} {Springer Proceedings in Mathematics \& Statistics}\ }\textbf
  {\bibinfo {volume} {282}},\ \bibinfo {pages} {569} (\bibinfo {year}
  {2019})}\BibitemShut {NoStop}%
\bibitem [{\citenamefont {Gallavotti}(2021)}]{Ga021a}%
  \BibitemOpen
  \bibfield  {author} {\bibinfo {author} {\bibfnamefont {G.}~\bibnamefont
  {Gallavotti}},\ }\href {\doibase 10.1007/s10955-021-02830-1} {\bibfield
  {journal} {\bibinfo  {journal} {Journal of Statistical Physics}\ }\textbf
  {\bibinfo {volume} {185}} (\bibinfo {year} {2021}),\
  10.1007/s10955-021-02830-1}\BibitemShut {NoStop}%
\bibitem [{\citenamefont {Shukla}\ \emph {et~al.}(2019)\citenamefont {Shukla},
  \citenamefont {Dubrulle}, \citenamefont {Nazarenko}, \citenamefont
  {Krstulovic},\ and\ \citenamefont {Thalabard}}]{SDNKT018}%
  \BibitemOpen
  \bibfield  {author} {\bibinfo {author} {\bibfnamefont {V.}~\bibnamefont
  {Shukla}}, \bibinfo {author} {\bibfnamefont {B.}~\bibnamefont {Dubrulle}},
  \bibinfo {author} {\bibfnamefont {S.}~\bibnamefont {Nazarenko}}, \bibinfo
  {author} {\bibfnamefont {G.}~\bibnamefont {Krstulovic}}, \ and\ \bibinfo
  {author} {\bibfnamefont {S.}~\bibnamefont {Thalabard}},\ }\href {\doibase
  10.1103/PhysRevE.100.043104} {\bibfield  {journal} {\bibinfo  {journal}
  {Phys. Rev. E}\ }\textbf {\bibinfo {volume} {100}},\ \bibinfo {pages}
  {043104} (\bibinfo {year} {2019})}\BibitemShut {NoStop}%
\bibitem [{\citenamefont {Jaccod}\ and\ \citenamefont
  {Chibbaro}(2021)}]{JC020}%
  \BibitemOpen
  \bibfield  {author} {\bibinfo {author} {\bibfnamefont {A.}~\bibnamefont
  {Jaccod}}\ and\ \bibinfo {author} {\bibfnamefont {S.}~\bibnamefont
  {Chibbaro}},\ }\href {\doibase 10.1103/PhysRevLett.127.194501} {\bibfield
  {journal} {\bibinfo  {journal} {Phys. Rev. Lett.}\ }\textbf {\bibinfo
  {volume} {127}},\ \bibinfo {pages} {194501} (\bibinfo {year}
  {2021})}\BibitemShut {NoStop}%
\bibitem [{\citenamefont {Fefferman}(2000)}]{Fe000}%
  \BibitemOpen
  \bibfield  {author} {\bibinfo {author} {\bibfnamefont {C.}~\bibnamefont
  {Fefferman}},\ }\href {\doibase
  https://www.researchgate.net/publication/281612631} {\emph {\bibinfo {title}
  {{Existence \& smoothness of the Navier–Stokes equation}}}},\ The
  millennium prize problems\ (\bibinfo  {publisher} {Clay Mathematics
  Institute},\ \bibinfo {address} {Cambridge, MA},\ \bibinfo {year} {2000})\
  pp.\ \bibinfo {pages} {57--67}\BibitemShut {NoStop}%
\bibitem [{\citenamefont {Maxwell}(2011)}]{Maxwell2011}%
  \BibitemOpen
  \bibfield  {author} {\bibinfo {author} {\bibfnamefont {J.~C.}\ \bibnamefont
  {Maxwell}},\ }\href {\doibase 10.1017/CBO9780511710377} {\emph {\bibinfo
  {title} {The Scientific Papers of James Clerk Maxwell}}},\ edited by\
  \bibinfo {editor} {\bibfnamefont {W.~D.}\ \bibnamefont {Niven}},\ \bibinfo
  {series} {Cambridge Library Collection - Physical Sciences}, Vol.~\bibinfo
  {volume} {2}\ (\bibinfo  {publisher} {Cambridge University Press},\ \bibinfo
  {year} {2011})\BibitemShut {NoStop}%
\bibitem [{\citenamefont {Gallavotti}(2006)}]{Ga006c}%
  \BibitemOpen
  \bibfield  {author} {\bibinfo {author} {\bibfnamefont {G.}~\bibnamefont
  {Gallavotti}},\ }\href {\doibase 10.1063/1.2372713} {\bibfield  {journal}
  {\bibinfo  {journal} {Chaos}\ }\textbf {\bibinfo {volume} {16}},\ \bibinfo
  {pages} {043114 (+6)} (\bibinfo {year} {2006})}\BibitemShut {NoStop}%
\bibitem [{\citenamefont {Alexakis}\ and\ \citenamefont
  {Brachet}(2020)}]{AB020}%
  \BibitemOpen
  \bibfield  {author} {\bibinfo {author} {\bibfnamefont {A.}~\bibnamefont
  {Alexakis}}\ and\ \bibinfo {author} {\bibfnamefont {M.}~\bibnamefont
  {Brachet}},\ }\href@noop {} {\bibfield  {journal} {\bibinfo  {journal}
  {Journal of Fluid Mechanics}\ }\textbf {\bibinfo {volume} {884}},\ \bibinfo
  {pages} {A33} (\bibinfo {year} {2020})}\BibitemShut {NoStop}%
\bibitem [{\citenamefont {Orszag}(1971)}]{Orszag1971}%
  \BibitemOpen
  \bibfield  {author} {\bibinfo {author} {\bibfnamefont {S.~A.}\ \bibnamefont
  {Orszag}},\ }\href@noop {} {\bibfield  {journal} {\bibinfo  {journal}
  {Journal of the Atmospheric sciences}\ }\textbf {\bibinfo {volume} {28}},\
  \bibinfo {pages} {1074} (\bibinfo {year} {1971})}\BibitemShut {NoStop}%
\bibitem [{\citenamefont {Pekurovsky}(2012)}]{P3DFFT}%
  \BibitemOpen
  \bibfield  {author} {\bibinfo {author} {\bibfnamefont {D.}~\bibnamefont
  {Pekurovsky}},\ }\href {\doibase 10.1137/11082748X} {\bibfield  {journal}
  {\bibinfo  {journal} {SIAM Journal on Scientific Computing}\ }\textbf
  {\bibinfo {volume} {34}},\ \bibinfo {pages} {C192} (\bibinfo {year}
  {2012})},\ \Eprint {http://arxiv.org/abs/https://doi.org/10.1137/11082748X}
  {https://doi.org/10.1137/11082748X} \BibitemShut {NoStop}%
\bibitem [{\citenamefont {Brun}\ and\ \citenamefont {Pumir}(2001)}]{Brun2001}%
  \BibitemOpen
  \bibfield  {author} {\bibinfo {author} {\bibfnamefont {C.}~\bibnamefont
  {Brun}}\ and\ \bibinfo {author} {\bibfnamefont {A.}~\bibnamefont {Pumir}},\
  }\href {\doibase 10.1103/PhysRevE.63.056313} {\bibfield  {journal} {\bibinfo
  {journal} {Phys. Rev. E}\ }\textbf {\bibinfo {volume} {63}},\ \bibinfo
  {pages} {056313} (\bibinfo {year} {2001})}\BibitemShut {NoStop}%
\bibitem [{\citenamefont {Leray}(1934)}]{Le934}%
  \BibitemOpen
  \bibfield  {author} {\bibinfo {author} {\bibfnamefont {J.}~\bibnamefont
  {Leray}},\ }\href@noop {} {\bibfield  {journal} {\bibinfo  {journal} {Acta
  Mathematica}\ }\textbf {\bibinfo {volume} {63}},\ \bibinfo {pages} {193}
  (\bibinfo {year} {1934})}\BibitemShut {NoStop}%
\bibitem [{\citenamefont {Caffarelli}\ \emph {et~al.}(1982)\citenamefont
  {Caffarelli}, \citenamefont {Kohn},\ and\ \citenamefont
  {Nirenberg}}]{CKN982}%
  \BibitemOpen
  \bibfield  {author} {\bibinfo {author} {\bibfnamefont {L.}~\bibnamefont
  {Caffarelli}}, \bibinfo {author} {\bibfnamefont {R.}~\bibnamefont {Kohn}}, \
  and\ \bibinfo {author} {\bibfnamefont {L.}~\bibnamefont {Nirenberg}},\ }\href
  {\doibase DOI: 10.1002/cpa.3160350604} {\bibfield  {journal} {\bibinfo
  {journal} {Communications on Pure and Applied Mathematics}\ }\textbf
  {\bibinfo {volume} {35}},\ \bibinfo {pages} {771} (\bibinfo {year}
  {1982})}\BibitemShut {NoStop}%
\bibitem [{\citenamefont {Gallavotti}(2005)}]{Ga002}%
  \BibitemOpen
  \bibfield  {author} {\bibinfo {author} {\bibfnamefont {G.}~\bibnamefont
  {Gallavotti}},\ }\href@noop {} {\emph {\bibinfo {title} {Foundations of Fluid
  Dynamics}}}\ (\bibinfo  {publisher} {(second printing) Springer Verlag},\
  \bibinfo {address} {Berlin},\ \bibinfo {year} {2005})\BibitemShut {NoStop}%
\bibitem [{\citenamefont {J.Serrin}(1962)}]{Se962}%
  \BibitemOpen
  \bibfield  {author} {\bibinfo {author} {\bibnamefont {J.Serrin}},\
  }\href@noop {} {\bibfield  {journal} {\bibinfo  {journal} {{Archive for
  Rational Mechanics and Analysis}}\ }\textbf {\bibinfo {volume} {9}},\
  \bibinfo {pages} {187} (\bibinfo {year} {1962})}\BibitemShut {NoStop}%
\bibitem [{\citenamefont {Kraichnan}(1975)}]{Kr975a}%
  \BibitemOpen
  \bibfield  {author} {\bibinfo {author} {\bibfnamefont {R.~H.}\ \bibnamefont
  {Kraichnan}},\ }\href@noop {} {\bibfield  {journal} {\bibinfo  {journal}
  {Advances in Mathematics}\ }\textbf {\bibinfo {volume} {16}},\ \bibinfo
  {pages} {305} (\bibinfo {year} {1975})}\BibitemShut {NoStop}%
\bibitem [{\citenamefont {Sobolev}(1963)}]{So963}%
  \BibitemOpen
  \bibfield  {author} {\bibinfo {author} {\bibfnamefont {S.}~\bibnamefont
  {Sobolev}},\ }\href@noop {} {\emph {\bibinfo {title} {{Applications of
  Functional analysis in Mathematical Phy\-sics}}}},\ \bibinfo {series}
  {Translations of the American Mathematical Society}, Vol.~\bibinfo {volume}
  {7}\ (\bibinfo  {publisher} {AMS},\ \bibinfo {address} {Providence},\
  \bibinfo {year} {1963})\BibitemShut {NoStop}%
\end{thebibliography}%
		
	\end{document}